\documentclass[sigconf,authorversion,nonacm]{acmart}
\AtBeginDocument{%
  }

\usepackage{soul}
\usepackage{bm}
\usepackage[utf8]{inputenc}
\usepackage{graphicx}

\usepackage{bbm}
\usepackage{multirow}
\usepackage{booktabs}
\usepackage{makecell}

\usepackage{graphicx}

\usepackage{natbib}
\usepackage{amsfonts}
\usepackage{soul}
\usepackage{xspace}
\usepackage{amsmath}
\usepackage{algorithm}
\usepackage{algpseudocode}
\usepackage{xcolor}     
\usepackage{float}
\usepackage{subcaption}

\newcommand{\ie}{\emph{i.e.,}\xspace}

\usepackage{enumitem}
\setlist[itemize]{itemsep=-0.25ex,leftmargin=2.5ex}
\setlist[enumerate]{itemsep=-0.25ex,leftmargin=2.5ex}

\usepackage{booktabs}

\renewcommand\hl[1]{#1}

\begin{document}

\title{Conversational Explanations: Discussing Explainable AI with Non-AI Experts}


\author{Tong Zhang}
\email{tong.zhang@ntu.edu.sg}
\affiliation{
  \institution{Nanyang Technological University}
  \country{Singapore}
}

\author{Mengao Zhang}
\email{zh0024ao@e.ntu.edu.sg}
\affiliation{
  \institution{Nanyang Technological University}
  \country{Singapore}
}

\author{Wei Yan Low}
\email{p180026@e.ntu.edu.sg}
\affiliation{
  \institution{Nanyang Technological University}
  \country{Singapore}
}

\author{X. Jessie Yang}
\email{xijyang@umich.edu}
\affiliation{
  \institution{University of Michigan}
  \city{Ann Arbor}
  \state{Michigan}
  \country{USA}
}

\author{Boyang Li}
\email{boyang.li@ntu.edu.sg}
\affiliation{
  \institution{Nanyang Technological University}
  \country{Singapore}
}

\renewcommand{\shortauthors}{Zhang et al.}

\begin{abstract}
\hl{Explainable AI (XAI) aims to provide insights into the decisions made by AI models.} To date, most XAI approaches provide only one-time, static explanations, which cannot cater to users' diverse knowledge levels and information needs.  Conversational explanations have been proposed as an effective method to customize XAI explanations. However, building conversational explanation systems is hindered by the scarcity of training data. Training with synthetic data faces two main challenges: lack of data diversity and hallucination in the generated data. To alleviate these issues, we introduce a repetition penalty to promote data diversity and exploit a hallucination detector to filter out untruthful synthetic conversation turns. We conducted both automatic and human evaluations on the proposed system, fEw-shot Multi-round ConvErsational Explanation (EMCEE). For automatic evaluation, EMCEE achieves relative improvements of 81.6\% in BLEU and 80.5\% in ROUGE compared to the baselines. EMCEE also mitigates the degeneration of data quality caused by training on synthetic data. In human evaluations ($N=60$), EMCEE outperforms baseline models and the control group in improving users' comprehension, acceptance, trust, and collaboration with static explanations by large margins. Through a fine-grained analysis of model responses, we further demonstrate that training on self-generated synthetic data improves the model’s ability to generate more truthful and understandable answers, leading to better user interactions. To the best of our knowledge, this is the first conversational explanation method that can answer free-form user questions following static explanations.

\end{abstract}



\maketitle

\section{Introduction}
Despite the high accuracy of deep neural networks (DNNs), it remains necessary for human domain experts to verify the DNN decisions and examine the reasoning process to prevent catastrophic failures in high-stake and mission-critical applications like healthcare, finance, and law enforcement.  \citep{Caruana2015health,Powles2017}. To this end, much research in recent years has been devoted to eXplainable Artificial Intelligence, or XAI (e.g., \citep{selvaraju2017grad, lundberg2017unified, chen2021hydra}). 

However, most current XAI techniques provide one-off, static explanations that are not customized to the user. As users differ in their knowledge levels, as well as tasks or goals that they try to accomplish, they will inherently have different information needs \citep{wang2021are, ehsan2021explainable, liao2020questioning, xin2023what}. Existing static XAI methods fail to address these diverse needs, leading to users' insufficient understanding of model behavior and undermining human-AI collaboration \citep{liao2020questioning, liao2021human, zhang2023may, nguyen2024allowing}. Indeed, recent studies found that the end users and domain experts with limited machine learning knowledge struggle to understand and use the XAI explanations \citep{ehsan2021explainable, wang2021are}.

Conversational explanations, or conversational XAI, have been suggested as a suitable solution for providing customized explanations to users \citep{liao2020questioning, feldhus2022mediators,lakkaraju2022rethinking, zhang2023may}.
\citet{lakkaraju2022rethinking} discovered that human decision-makers have a strong preference for explanations in the form of natural language dialogue. They argued that conversational explanations can provide personalized responses and relevant information based on users' conversational histories. \citet{zhang2023may} showed that answering users' follow-up questions after providing static explanations significantly improves their comprehension, acceptance, trust, and collaborative decision-making with AI. While the need for conversational XAI has been recognized, building such systems is hindered by data scarcity, partially due to the difficulty of collecting high-quality conversations about AI explanations. As far as we are aware, there is only one dataset of 60 conversations focused on two types of static explanations \citep{zhang2023may}. To date, existing conversational explanations rely on human-authored templates, which can only handle a limited and predefined range of user questions \citep{slack2023explaining, shen2023convxai}.

To handle data scarcity, we propose a novel method in this work to develop conversational explanations by training large vision language models (VLMs) on synthetic conversations. However, training with synthetic data encounters two primary challenges: the lack of data diversity in self-generated data \citep{schwarz2021frequency, shumailov2024curse,briesch2023large}, and the hallucinations generated by VLMs \citep{lee2022factuality, ji2023survey, dai-etal-2023-plausible, zheng2023does, berglund2024the}. 
The first challenge, lack of data diversity, arises as generative models tend to overrepresent high-frequency content~\citep{schwarz2021frequency, shumailov2024curse,briesch2023large} and suppress the tails of the data distribution. To alleviate this issue, we introduce a repetition penalty that reduces the frequency of tokens existing in previously generated conversations. The other obstacle is the hallucination in generated conversations. VLMs often suffer from generating untruthful information, referred to as hallucination~\citep{lee2022factuality, ji2023survey, dai-etal-2023-plausible, zheng2023does, berglund2024the}. To mitigate the hallucinated, factually incorrect answers, we trained a hallucination detector to filter out such conversation turns after data generation. To train the detector, we collected a hallucination dataset of 750 factual and 750 incorrect statements about basic machine learning and XAI methods. 

We conducted both automatic and human evaluations on the proposed system, fEw-shot Multi-round ConvErsational Explanation (EMCEE), to assess its performance. The automatic evaluation is conducted on the only existing conversational explanation dataset~\citep{zhang2023may}. We assessed the performance of different conversational XAI systems by measuring the word overlap between generated responses and ground truth texts.

For the human evaluation, we evaluated how different conversational XAI systems assist users in understanding static explanations of image classification models, improving acceptance and trust in XAI methods, and choosing the best AI models using only the explanations. We recruited a total of 60 participants and randomly divided them into three groups of equal size. One group interacted with our EMCEE model regarding static explanations, another group engaged with the baseline LLaVa-1.5 model, and the control group independently reviewed materials about the static explanations. Before and after the conversation or reading session, we measured their objective understanding and subjective perceptions of the provided static explanations. Based on the results, we estimated the effectiveness of the different conversational explanation systems in improving users' comprehension and usage of explanations.

Empirical results showed that our EMCEE outperforms the baseline LLaVa-1.5 model in both automatic and human evaluations by a large margin. In the automatic evaluation, EMCEE achieved relative improvements of 81.6\% in BLEU and 80.5\% in ROUGE compared to the baseline. While repeated training on self-generated data often leads to reduced diversity and quality \citep{briesch2023large}, we showed that the proposed repetition penalty and hallucination detection can slow down the data degeneracy in training with synthetic data. 
In the human evaluation, participants who interacted with EMCEE reported a better understanding of static explanations, felt that the explanations enhanced their experience with AI models, were more inclined to use explanations in the future, trusted the explanations more, and demonstrated that they could collaborate better with AI systems using the explanations.

To further investigate how training on self-generated synthetic data enhances user interactions with the conversational XAI system, we conducted a fine-grained analysis of model responses to different types of user questions. We manually classified the questions asked during the human evaluation into three categories: generic AI/XAI questions, questions related to the provided explanations, and extended questions. We sampled 10 questions from each category for both the baseline and EMCEE models. Three well-educated annotators rated the responses on factual correctness and understandability. Results showed that EMCEE consistently provides more accurate and truthful answers across all question types compared to the baseline. The improvement in factual correctness highlights the effectiveness of the hallucination detector in filtering out incorrect statements from the synthetic data and reducing model hallucinations. In terms of understandability, EMCEE outperforms the baseline, particularly for questions related to the provided explanations. This suggests that training on synthetic conversations helps EMCEE better grasp the conversational context of explanations, leading to more understandable responses for users.

Our contributions can be summarized as follows.
\begin{itemize}[topsep=0.1em]
    \item To the best of our knowledge, we propose the first conversational explanation system that can answer free-form follow-up questions after providing static explanations to users.
    \item We introduce a novel method to train conversational explanation systems on self-generated synthetic data. To enhance data quality, we propose a repetition penalty to boost data diversity and a hallucination detector to reduce erroneous information in synthetic data.
    \item We validate the effectiveness of our conversation explanation system, EMCEE, through both automatic and human evaluation~($N=60$). Results show that EMCEE significantly outperforms baseline models in helping non-AI experts understand and utilize AI explanations.
    \item We analyze model responses to user questions and demonstrate that training on self-generated synthetic data improves the model’s ability to generate more truthful and understandable responses, leading to enhanced user interactions with the system.
\end{itemize}
\section{Related Work}

\subsection{Static XAI}
Explainable Artificial Intelligence (XAI) refers to techniques that explain the learning process or the predictions of AI~\citep{yang2019evaluating}. Most existing techniques are static XAI, which provides a one-time explanation with no capability for further user interaction. These techniques can be broadly divided into two categories: self-explanatory models and post-hoc methods. Self-explanatory models are inherently transparent, offering clarity in their decision-making processes~\citep{lakkaraju2016interpretable, rudzinski2016multi,yang2017scalable, jain2019attention, wiegreffe-pinter-2019-attention}. 
The majority of recent XAI methods are post-hoc XAI methods, applied to already developed models that lack inherent transparency~\citep{selvaraju2017grad, ribeiro2016should, chen2021hydra, adadi2018peeking, bodria2021benchmarking}. 
There are two main groups of methods in post-hoc XAI, i.e., feature attribution methods and example-based methods.

\noindent \textbf{Feature Attribution.} Feature attribution methods explain model predictions by investigating the importance of input features to final predictions \citep{adadi2018peeking, danilevsky-etal-2020-survey}. There are two main types of feature attribution methods, gradient-based methods~\citep{cortez2013using, sundararajan2017axiomatic, selvaraju2017grad, simonyan2013deep, lundberg2017unified, wang2024gradient, kokalj2021bertShapley, li2016visualizing, wang2024gradient} and surrogate methods \citep{ribeiro2016should,hu2018locally, alvarez2017causal, liu2018interpretation, shih2018symbolic, ignatiev2019abduction}. Gradient-based methods employ gradients to evaluate the contribution of a model input on the model output. 
Surrogate methods leverage a simple and inherently interpretable model, such as a linear model, to locally approximate the behavior of the complex neural network. 

\noindent \textbf{Example-based Methods.} Example-based methods explain AI predictions by identifying a selection of data instances~\citep{adadi2018peeking, danilevsky-etal-2020-survey, pcnn2024nguyen}. These instances may be training data points with the most influence on the parameters of a prediction model~\citep{chen2021hydra, guo2021fastif}, counterfactual examples that alter predictions with minimal changes to inputs \citep{ wachter2017counterfactual,mothilal2020explaining, yin-neubig-2022-interpreting, ye2021connecting, ross2021explaining, wu2021polyjuice}, or prototypes that contain semantically similar parts to input instances \citep{croce2019auditing,jeyakumar2020can,kim2016examples}.

\hl{In this work, we focus primarily on feature attribution methods, as they directly highlight the importance of input features, making the decision-making process of models more intuitive for laypeople {\citep{kim2023help}}. Specifically, we select Grad-CAM, Integrated Gradients, and SHAP from gradient-based methods, as well as LIME from surrogate methods, to evaluate the effectiveness of different conversational evaluation systems.}

\subsection{Conversational XAI}
Human-Computer Interaction (HCI) researchers have recently proposed that XAI methods should involve conversation, aligning with the natural way humans explain to each other. Specifically, \citet{lombrozo2006structure} argues that explanations emerge from a conversational interaction between an explainer and an explainee. Similarly, \citet{miller2019explanation} emphasizes that explanations should include an interactive communication process, where the explainer provides the necessary information for the explainee to understand the causes of an event through dialogue. Building on this perspective of human explanations, recent works have introduced the concept of "explainability as dialogue," aiming to make explanations more accessible to a wide range of non-expert users \citep{liao2020questioning, feldhus2022mediators, lakkaraju2022rethinking}.

Despite much exploration of the role of conversation in explainability, the practical development of conversational XAI is still in its early stages, with limited methods available so far. \citet{shen2023convxai} applied conversational explanations to scientific writing tasks, finding improvements in productivity and sentence quality. Likewise, \citet{slack2023explaining} designed dialogue systems that help users better understand machine learning models in tasks like diabetes prediction, rearrest prediction, and loan default prediction. However, these systems rely on template-generated conversations and can only handle a limited set of predefined queries. Our work represents the first system capable of delivering free-form explanatory conversations about static explanations.

\subsection{Training with Synthetic Data}
The exceptional performance of Large Language Models (LLMs) and Vision Language Models (VLMs) in generating human-like text has encouraged researchers to explore their use as training data generators \citep{meng2022generating, ye-etal-2022-zerogen, guo2024generative, gao2023selfguided,meng2023tuning, ye2022progen}. For example, SuperGen \citep{meng2022generating} uses LLMs conditioned on label-descriptive prompts to generate training data for text classification tasks. 
FewGen \citep{meng2023tuning} finetune an LLM on few-shot samples and uses it to generate synthetic data for seven classification tasks in the GLUE benchmark. While LLMs and VLMs have shown promise in generating human-like texts, they still face the challenge of producing noisy and low-quality synthetic data. This may lead to decreased performance or perpetuated biases in the model trained on the data \citep{schwarz2021frequency, zhang2024trade,kirk2021bias, esiobu2023robbie, lee2022factuality, ji2023survey}.

To mitigate the detrimental effects of noisy and low-quality synthetic data from LLMs and VLMs, several methods have been proposed \citep{gao2023selfguided, guo2024generative, meng2023tuning, ye2022progen}. For example, ProGen \citep{ye2022progen} adjusts the weight of generated data points with regard to its influence on the validation loss, using influence function \citep{koh2017understanding}.  However, these strategies have primarily focused on generating data for classification tasks and on training small-scale task-specific models. Techniques such as applying the influence function to weigh data points are effective for smaller models. They present challenges and require a special design when adapted to LLMs \citep{grosse2023studying}.

In our work, we apply data generation to conversational explanations and utilize generated data to train the original VLM. We improved the quality of the generated data and significantly slowed down model degeneracy after multiple generation-training iterations (see \S \ref{sec:slowed-degeneration}). 

\begin{figure*}
    \centering
    \includegraphics[width=0.9\textwidth]{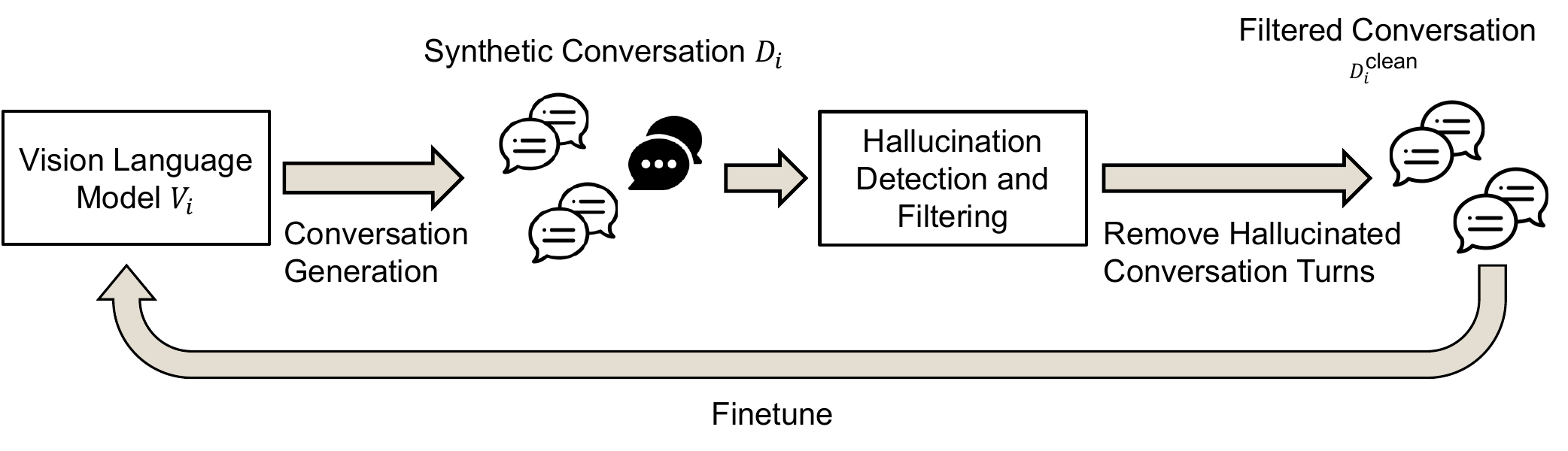}
    \caption{The Overall Workflow of EMCEE. $V_i$ denotes the VLM and $D_i$ denotes the synthetic conversation data in the $i$-th iteration. Starting from a pretrained VLM $V_1$, we first generate diverse synthetic conversations $D_1$ with the repetition penalty. Next, we use a hallucination detector to clean synthetic data, producing cleaned data $D_1^{\text{\,clean}}$. We then finetune the VLM on $D_1^{\text{\,clean}}$, which creates $V_2$, and this process repeats.  
    }
    \label{fig:model_structure}
    \vspace{-10pt}
\end{figure*}

\section{Methodology}
The overall workflow of EMCEE is illustrated as Figure \ref{fig:model_structure} and outlined in Algorithm \ref{alg:algorithm}. 
Starting from a pretrained VLM~$V_1$, we generated a set of synthetic conversations $D_1$, while using the repetition penalty to encourage data diversity. Each conversation contains multiple turns, denoted as $\langle (\bm{x}_1, \bm{y}_1), (\bm{x}_2, \bm{y}_2), \ldots \rangle$, where the human turn is $\bm{x}_i$ and the machine response is $\bm{y}_i$. Then, we applied a hallucination detector $f_h$, which filters out hallucinated conversation turns. That is, if we detect hallucination from the machine response (\ie $f_h(\bm{y}_i) = 1$), $(\bm{x}_i, \bm{y}_i)$ is removed from the conversation. This process yields cleaned data $D_1^{\text{\,clean}}$. Afterwards, we finetuned the VLM on $D_1^{\text{\,clean}}$, leading to the next VLM $V_2$, from which we start another round of generation-filtering-finetuning. This process is repeated multiple times, without reusing any synthetic data from previous rounds. 

We designed a prompt that is used across all stages, \ie data generation, model fine-tuning, and model inference. The prompt includes an instruction, background information about the AI model and XAI method, and several demonstration conversations. The instruction specifies the purpose of the conversation, which is to enhance user comprehension of static explanations. The background information includes details about the prediction task, the machine learning model, the XAI technique, and an example explanation. Details of the prompts are in Appendix \ref{sec:prompt_descrip}.

The number of demonstration conversations utilized varies in different stages. During data generation and model finetuning, we randomly chose zero or one demonstration and kept it consistent for each mini-batch. During model inference and evaluation, the number of demonstrations ranged between zero and three.

\subsection{Repetition Penalty}

The repetition penalty encourages the VLM to generate more diverse conversations by discounting the logits of tokens seen in previous conversation turns. Specifically, given the logits $z_i$ for each token $i$ in the vocabulary, the probability $p_i$ of predicting token $i$ is computed as,
\begin{equation}
        p_i = \frac{\exp(z_i/(T + \theta \cdot \mathbbm{1}(i \in G)))}{\sum_j \exp(z_j/(T + \theta \cdot \mathbbm{1}(j \in G)))},
\end{equation}
where $T$ is the temperature. $\theta$ is the ratio of the repetition penalty. $G$ is the set of words existing in generated conversations in the current round, and $\mathbbm{1}$ is an indicator function. When the token $i$ exists in $G$, $\mathbbm{1}(i \in G)$ is 1, otherwise, $\mathbbm{1}(i \in G)$ is 0.

\begin{algorithm}[!t]
\caption{EMCEE}
\label{alg:algorithm}
\begin{algorithmic}[1] 
\Statex \hspace{-\algorithmicindent} \textbf{Input}: a pretrained VLM $V_1$; a hallucination detector $f_h$, $f_h(\bm{y}) = 1$ if $\bm{y}$ is deemed  hallucination; number of conversations \Statex \hspace{-\algorithmicindent} to generate per round $N$; maximum number of rounds $R$.
\Statex \hspace{-\algorithmicindent} \textbf{Output}: a finetuned model $V_R$
\For{$r$ \textbf{in} $1...R$}
    \State $\mathcal{D}_r$ $\leftarrow$ generate $N$ conversations from  $V_{r}$;
    \State $D_i^{\text{\,clean}}$ $\leftarrow  \{(\bm{x}, \bm{y})\in D_r \mid f_h(\bm{y}) \neq 1\} $;
    \State $V_{r+1}$ $\leftarrow$ finetune $V_{r}$ on $D_i^{\text{\,clean}}$;
\EndFor
\end{algorithmic}
\end{algorithm}

\subsection{Hallucination Detection and Filtering}
VLMs often generate convincing but factually incorrect statements, especially when answering questions that require reasoning and logical deduction \citep{lee2022factuality, ji2023survey, dai-etal-2023-plausible, zheng2023does, berglund2024the}. Conversational explanations are mainly about explaining the causal relationship between static explanations and AI predictions, which involves significant reasoning. Therefore, hallucination is a major concern in this use case.

To reduce hallucination, we integrated a hallucination detector into the training process, which identifies and removes hallucinated conversation turns. To train the detector, we constructed a dataset comprising 1,500 sentences about machine learning and XAI methods. The dataset is balanced, containing 750 factually correct sentences and~750 factually incorrect ones. It includes 500 sentences on general machine learning knowledge, sourced from students studying machine learning. Among 500 sentences, 250 sentences were picked from the class notes by students. After that, the students were asked to create 250 incorrect sentences based on the correct ones. The remaining 1,000 sentences are about XAI knowledge; we used GPT-4-turbo-2024-04-09 to generate 500 factually correct sentences about XAI and subsequently alter them to be incorrect. All generated sentences have been rigorously validated by XAI experts. To prevent data duplication, we manually removed all duplicated sentences.
Example sentences from this dataset are displayed in Table \ref{tab:hallucination_dataset_example}. We used 80\% of the dataset for training the detector, with the remaining 20\% reserved for validation and testing.

\begin{table*}[ht]
    \begin{tabular}
    {p{12cm} c} 
    \hline
    \textbf{Sentence} & \textbf{Label}\\
    \hline
    When the amount of data stays the same, the more parameters, the more difficult to estimate the parameters accurately. & 0\\
    When the amount of data stays the same, increasing the number of parameters can improve the accuracy of their estimates. & 1\\
    \hline
    XAI is less important in systems where decisions are not critical. & 0\\
    XAI is only relevant in non-critical systems. & 1\\
    \hline
    Grad-CAM can be applied to any convolutional layer of a network, not just the final layer. & 0 \\
    Grad-CAM is restricted to analyzing the input and output layers of a network. & 1\\
    \hline
    LIME can explain any machine learning model as long as it can probe the model with perturbed inputs. & 0\\
    LIME can only explain models that are specifically designed to work with its framework. & 1\\
    \hline
    The path taken from baseline to input in Integrated Gradients is typically linear. & 0\\
    The path taken is randomly generated in each run of Integrated Gradients. & 1\\
    \hline
    SHAP values can be computed for any data point in the dataset, providing versatile insights. & 0\\
    SHAP values can only be computed for a limited set of predefined data points. & 1\\
    \hline
    \end{tabular}
    \caption{Examples of sentences with labels in our hallucination dataset. Label 0 means the sentence is factually correct; label 1 means the sentence is factually incorrect.}
    \label{tab:hallucination_dataset_example}
\end{table*}

\subsection{Implementation}
We used LLaVa-1.5 \citep{liu2023llava, liu2023improvedllava} as our base vision language model. LLaVa-1.5 is an end-to-end trained large multimodal model that combines a vision encoder and an LLM for general-purpose visual and language understanding. We chose \mbox{LLaVa-1.5} for its high performance in answering scientific questions and proficiency in visual chat scenarios \citep{liu2023llava, liu2023improvedllava}.

For the data generation process, the number of generated conversations $N$ at each round is set to~2000, with 500 conversations for each static explanation method. \hl{The number of training iterations of the generation network is empirically tuned on the validation set and set to five.} The temperature is set to 1.2 to encourage diverse generations while maintaining coherence. The repetition penalty ratio is set to~1.1. For finetuning LLaVa-1.5, we used LoRA \citep{hu2021lora} to only finetune the language model while keeping the vision encoder and projector frozen. The rank of the LoRA parameter is set to 128, the batch size is 32, and the learning rate is $2\times10^{-4}$ with cosine annealing. In each generation-filtering-finetuning round, we finetuned the LLaVa-1.5 for 3 epochs. Finally, for the hallucination detector, we trained a Bert-base model \citep{devlin2019bert} using the SGD optimizer with a learning rate of 0.01, batch size of 16, and weight decay for 100 epochs. The hallucination detector achieved an accuracy of 79.5\% on the held-out test set.

\section{Evaluation Methodology}
In this section, we present the evaluation methodology used to assess the performance of our proposed EMCEE model. We employed two evaluation methods: automatic and human evaluations. The automatic evaluation is crucial for objectively measuring the model’s ability to generate responses that align with ground truth explanations, using established metrics. Since our conversational XAI system is designed to help users better understand and utilize static explanations, human evaluation is necessary to assess its real-world impact. We examined the system’s effectiveness by observing participants' comprehension, acceptance, trust in the static explanations, and their ability to collaborate with these explanations, both before and after interacting with the system. 

\subsection{Automatic Evaluation Metrics and Dataset}
For automatic evaluations, we conducted few-shot evaluations with zero to three demonstrations. We leverage BLEU~\citep{papineni-etal-2002-bleu} and ROUGE \citep{lin-och-2004-automatic} scores to measure word overlaps between generated response text and ground truth text. 
Higher BLEU and ROUGE scores indicate better alignment between the generated and human-written texts, reflecting the model’s ability to produce more accurate and contextually appropriate outputs.
These two metrics are commonly used in natural language processing (NLP) evaluation, as they are easy to compute and comparable across different papers. 

We conducted our automatic evaluation using the only existing dataset of human-human conversational XAI interactions, which was collected in previous work by \citet{zhang2023may}. This dataset was gathered using a Wizard-of-Oz (WoZ) setting \citep{kelly1984wizard-of-oz}. Participants interacted with what they believed was an autonomous dialogue system, which was actually operated by a human expert in machine learning and XAI. The dataset includes 30 conversations on the LIME method and another 30 on the Grad-CAM method. On average, each conversation contains 27.4 utterances, with each utterance averaging 14.4 words. Due to its small size, we did not use this dataset for training. We employed one conversation per static explanation method (LIME and Grad-CAM) as a demonstration in the data generation prompt and six conversations for demonstrations in the few-shot evaluation. In the remaining 52 conversations, 10 conversations were used for validation and 42 were used for testing.

Although BLEU and ROUGE are useful and widely used, they have limitations. High n-gram overlap with human-written references does not necessarily guarantee that users can understand the generated responses or that the responses are coherent within the conversation. Therefore, we also conducted human evaluations to assess the effectiveness of different conversational models in helping users understand, accept, and trust static explanations.

\subsection{Human Evaluation Protocol}
For the human evaluation, we evaluated the effect of different conversational XAI methods by observing participants' objective understanding and subjective perception of static explanations, before and after interacting with different conversational XAI methods. Our study has received approval from our Institutional Review Board (\#IRB-2023-254). 

\begin{table}[t]
    \caption{Academic disciplines of our participants and the number of participants in each group. There are 60 participants from 4 different discipline groups.}
    \centering
    \begin{tabular}{l c}
        \toprule[1pt]
        \textbf{Academic Discipline} & \textbf{Number of Participants} \\
        \hline
        Business & 14\\
        Engineering & 10\\
        Humanities & 18\\
        Science & 18\\
        \bottomrule[1pt]
    \end{tabular}
    \label{tab:user_domain}
\end{table}

\subsubsection{Participants}
We recruited 60 participants for our study. All were 21 years old or older, fluent in English, and had not been involved in research about XAI previously. 
We recruited our participants in two ways: by posting advertisements on an online forum and by emailing students and staff across various departments and schools.
To ensure diversity, participants came from a broad range of disciplines. For ease of reporting, we categorize their disciplines into four groups:
\begin{itemize}
    \item Business, including Business and Accountancy.
    \item Engineering, including Civil and Environmental Engineering, Electrical and Electronics Engineering, Chemical Engineering and Biotechnology
    \item Humanities, including Psychology, Economics, Communication Studies, Linguistics and Multilingual Studies, and Sociology.
    \item Science, including Biology, Chemistry, Sport Science \& Management, and Physics.
\end{itemize}
Table \ref{tab:user_domain} shows statistics of the academic disciplines that the participants enrolled in.

\subsubsection{Experimental Task}
We focused on the image classification task on the ImageNet dataset and trained three classification models with different top-1 classification accuracies: Swin Transformer (84.1\%), VGG-16 (71.6\%), and AlexNet (56.5\%). We chose image classification because it requires minimal domain-specific expertise, making it well-suited for crowdsourcing among participants from diverse domains. 
To generate explanations for model predictions, we adopted four feature attribution explanation methods: LIME~\citep{ribeiro2016should}, Grad-CAM \citep{selvaraju2017grad}, Integrated Gradients \citep{sundararajan2017axiomatic}, and SHAP~\citep{lundberg2017unified}. For a more comprehensive evaluation, we extended the two XAI methods used in automatic evaluation to these four attribution explanation methods. The focus is on feature attribution as we believe the relationship between input features and model predictions is more intuitive to understand for laypeople than, for example, data attribution~\citep{kim2023help}.

\subsubsection{Experimental Interface} 
\label{sec_interface}
\begin{figure*}[t]
    \centering
    \includegraphics[width=\textwidth]{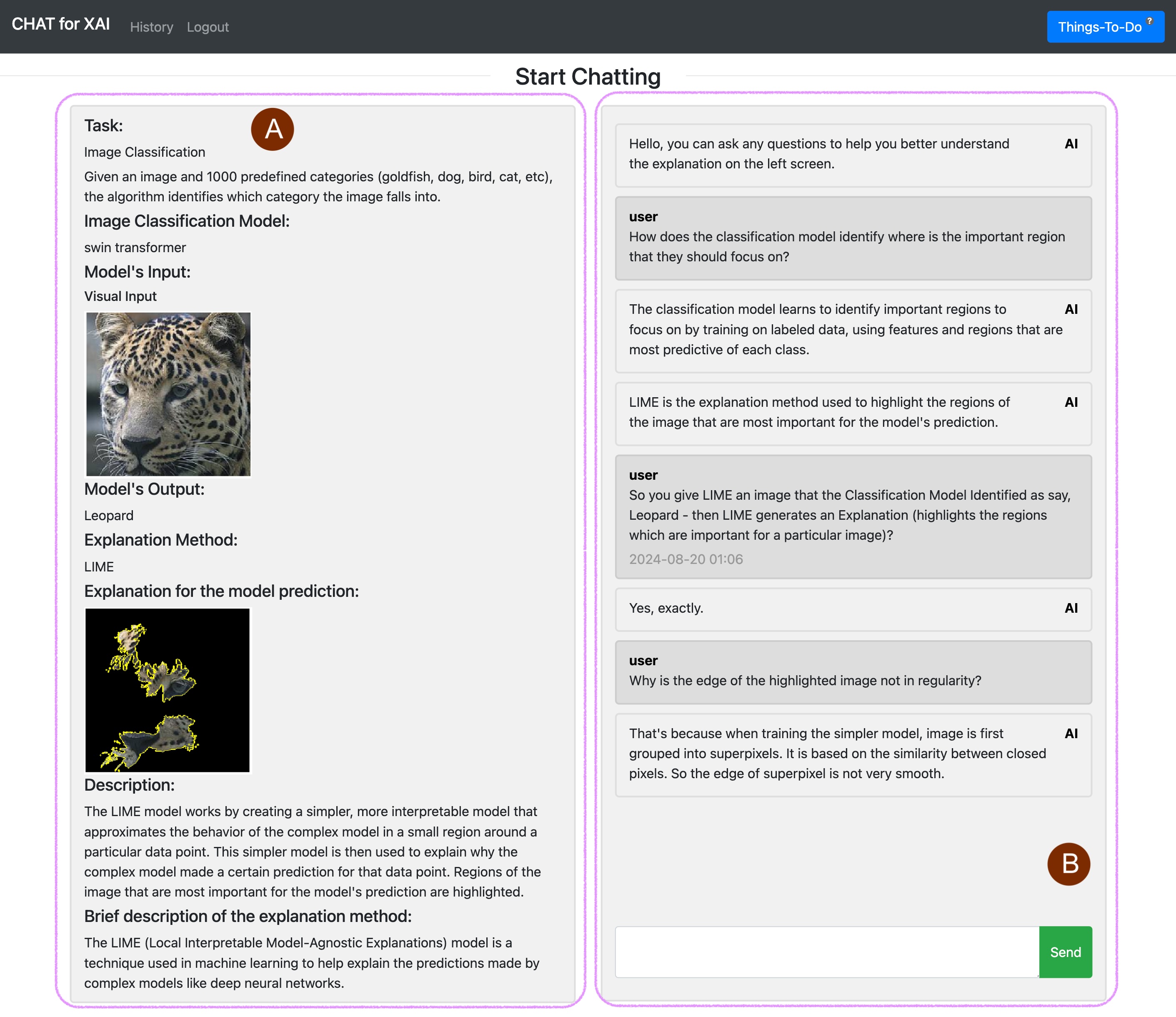}
    \caption{The interface where users discuss static explanations with a conversational agent. \textit{Part A}: Information about static explanations, including a task description, a description of the prediction model, a model input, a model output, an explanation generated by the explanation model, and a description of the explanation. \textit{Part B}: A chatbox where users converse with a conversational agent to clarify the explanation.}
    \label{fig:chat_page}
\end{figure*}

Our study was conducted on a web-based platform allowing participants to complete the entire procedure remotely. This platform ensures that all communication between users and conversational agents is text-based and recorded.
Figure \ref{fig:chat_page} displays an example screenshot of the interface. There are two sections on the page. On the left (Figure \ref{fig:chat_page} Part A), participants see a task description, a description of the prediction model, a model input, a model output, an explanation generated by the explanation model, and a description of the explanation. On the right within the chatbox (Figure \ref{fig:chat_page} Part B), participants engage in a text-based conversation with the agent to clarify the provided explanation. Participants can ask any questions or provide comments related to the explanation on the left. In the control group, we replaced the chatbox with a 15-minute timer. Once the timer reached zero, participants were allowed to proceed to the post-measurements.

\subsubsection{Experimental Design}
There are two independent variables and two categories of dependent variables. The first independent variable is the explanation method: LIME, Grad-CAM, Integrated Gradients, or SHAP. The second independent variable is the participant's group: participants either have a conversation with our EMCEE, a conversation with the baseline LLaVA-1.5, or read the static explanations. We measure participants' objective understanding and subjective perceptions of explanations before and after conversations or readings. Two sets of dependent variables are collected in the experiment: the model selection accuracy and the self-reported perception scores.

Previous work {\citep{ehsan2021explainable}} indicates that participants' prior knowledge of AI may influence their perceptions of explanations, potentially introducing a confounding factor. To mitigate any potential confounding variable, participant assignments to the three groups are completely random. Additionally, the results in Section~{\ref{sec:human_evaluation_result}} showed no significant difference (p = 0.44) amongst self-reported pre-conversation understanding of the three groups.

\subsubsection{Measurement of Users' Objective Understanding -- Selection of Classification Models}
We assessed users' understanding of static explanations by measuring their performance in a model selection task. Model selection is a fundamental task for machine learning practitioners \citep{anderson2004model}. Specifically, participants were presented with 5 input images, on which the three classification models make identical decisions. The only differences between these models are their explanations. Participants must choose the model that they believe will perform most accurately on unobserved test data. Hence, to make the correct selection, the participants must understand the explanations. We measured participants’ objective understanding of static explanations by their accuracy in selecting the correct model. The complete set of images used for LIME, Grad-CAM, Integrated Gradients, and SHAP is detailed in Appendix \ref{appendix_sec:objective}. 

We observe that static explanations do not always faithfully reflect the actual workings of classification models~\citep{Adebayo2018SanityChecks, Kindermans2019, jacovi2020towards} and do not always contain actionable information for model selection. In our study, model selection is used to determine whether users can comprehend static explanations \emph{when} the explanations do have actionable information for selection, rather than assessing the explanations themselves. For this, we chose images that models with high accuracy can provide more reasonable explanations. An explanation is deemed more reasonable when it highlights features that are unique to the predicted class and avoids irrelevant features. A good model should have explanations that rely on multiple types of discriminative features, making the model more robust.  Consequently, the model makes the correct decision even if some discriminative features are absent or occluded. Hence, this approach allows users to easily pick the best classification models if they understand the static explanations well. 

\hl{Other objective measurements to assess user understanding \mbox{\citep{wang2021are, cheng2019explaining}} are not suitable for our study. These methods measure users' accuracy on the same prediction tasks as the classifier, such as student admission {\citep{cheng2019explaining}} or recidivism prediction {\citep{wang2021are}}. Their main idea is to compare users' predictive accuracy with and without explanations. However, the current work focuses on the task of image classification, a task users can easily perform without any explanation. Even when the user cannot make correct predictions, explanations highlighting relevant portions of the image is unlikely to provide any assistance. Instead, we used model selection as the objective measurement. To ensure that the correct model selection reflects users' understanding of XAI, we carefully curated a set of images so that users could only make the correct choice if they comprehended the explanations provided.}

\subsubsection{Measurements of Users' Subjective Perception}
We also measured participants’ self-reported perception of the static explanations, including their comprehension, acceptance, and trust. There are a total of 13 questions. All questions utilize a 7-point
Likert scale for responses. The full list of the questions is in Appendix \ref{appendix_sec:subjective}. 

\begin{itemize}
    \item Comprehension \citep{hoffman2018metrics, cheng2019explaining}: Participants' subjective perceptions of their understanding of explanations. It serves as a supplement to objective assessments, offering a more comprehensive view of how well participants understand static explanations.
    \item Perceived Usefulness \citep{davis1989perceived, davis1989user, diop2019extension}: The degree to which participants feel that the explanations enhance their experience with deep learning models. Together with \textit{perceived ease of use} and \textit{behavioral intention}, these three factors measure participants' acceptance of static explanations. They are derived from the Technology Acceptance Model (TAM) \citep{davis1989perceived, davis1989user, diop2019extension},  a widely applied theory for understanding individual acceptance and usage of information systems. Investigating users' acceptance of the explanations is very important, as the explanations are intended for end-users.
    \item Perceived Ease of Use \citep{davis1989perceived, davis1989user, diop2019extension}: 
    Participants' judgment of the simplicity and clarity of the explanations.
    \item Behavioral Intention \citep{davis1989perceived, davis1989user, diop2019extension}: The tendency of participants to utilize the explanation information in the future.
    \item Trust \citep{muir1996trust, Tita2022Systematic}: Participants' confidence in the reliability of the explanation methods to perform as intended. Trust has been recognized as a key factor in human-AI collaboration, as it influences how much humans rely on AI models, thus directly affecting the effectiveness of the human-AI team \citep{sebo2020influence, seaborn2021voice, doshi2017towards, vorm2022Integrating, Andrew2023Explainable, Yang:2017:EEU:2909824.3020230, Guo2020}.
\end{itemize}

\subsubsection{Experimental Procedure} 
Before participating in the study, participants signed an informed consent form that details the study’s objectives and procedures. The form also explained the compensation and ensured both anonymity and confidentiality of the collected data. After signing, participants received an email with instructions to access the study platform. Once logged in, a pop-up window provided a brief overview of the tasks. Participants then began with pre-experiment measurements for their objective understanding and subjective perceptions of static explanations. Objective understanding was assessed by letting participants choose the most accurate of three classification models on unseen test data, using 5 explanation examples. The subjective perception was measured through 13 self-reporting questions. These questions probed participants' perceived comprehension, acceptance, and trust in the explanations. 

\hl{After these initial measurements, we randomly assigned participants into three groups of equal size. One-third engaged in an online text-based conversation with our EMCEE model to ask questions and clarify doubts. Another third conversed with the baseline LLaVA-1.5 model. The remaining third, serving as the control group, spent 15 minutes reviewing the static explanations independently. The duration matched the average time spent in conversations by the other two groups. The information provided to participants at this stage is displayed in Figure {\ref{fig:chat_page}}. Since this phase focuses on explaining XAI methods to users rather than testing their understanding, explanations for just one classifier were sufficient. We used the Swin Transformer as the classifier due to its higher classification accuracy.}

After the conversation or reading session, participants repeated the same measurements of objective understanding and subjective perceptions as in the pre-session phase. All results and conversation records are documented. Upon completing the study, participants received a \$10 reward.

\section{Results \& Discussion}
This section presents the experimental results from both automatic and human evaluations of the baseline and our EMCEE model. For the automatic evaluation, we report results on an existing dataset of human-human conversational XAI interactions and include an ablation study to show the effectiveness of different components in our method. For the human evaluation, we present results on participants' objective understanding and subjective perceptions of static explanations, measured before and after different conditions. We also analyze the collected conversations and provide insights into why our system can improve users' understanding, acceptance, and trust in static explanations.

\begin{table*}[htbp]
\centering
\caption{Automatic Evaluation of pretrained LLaVa-1.5 and our model. We prompt models with 0 to 3 example conversations.}
\label{tab:automatic_results}
\resizebox{\linewidth}{!}{
\begin{tabular}{cccccccccc}
\toprule
         Methods & Shot Num & BLEU-1 & BLEU-2 & BLEU-3 & BLEU-4 & ROUGE-1 & ROUGE-2 & ROUGE-3 & ROUGE-L \\
\midrule
\multirow{4}*{LLaVa-1.5} 
            & 0 & 0.1328 & 0.0534 & 0.0235 & 0.0103 & 0.3150 & 0.0595 & 0.0179 & 0.2507 \\
            & 1 & 0.1447 & 0.0680 & 0.0361 & 0.0196 & 0.2823 & 0.0823 & 0.0374 & 0.2324 \\
            & 2 & \underline{0.2160} & \underline{0.1329} & \underline{0.0985} & \underline{0.0813} & \underline{0.3365} & \underline{0.1469} & \underline{0.1014} & \underline{0.2883} \\
            & 3 & 0.1979 & 0.1265 & 0.0854 & 0.0687 & 0.3153 & 0.1339 & 0.0839 & 0.2709 \\
\hline
            \multirow{4}*{\makecell{EMCEE\\(Ours)}} 
            & 0 & 0.2394 & 0.1659 & 0.1270 & 0.1055 & 0.3918 & 0.2295 & 0.1794 & 0.3418 \\
            & 1 & 0.2895 & 0.2186 & 0.1826 & 0.1618 & 0.4513 & 0.2854 & 0.2391 & 0.4006 \\
            & 2 & \textbf{0.3056} & \textbf{0.2336} & \textbf{0.1945} & \textbf{0.1721} & \textbf{0.4629} & \textbf{0.2964} & \textbf{0.2454} & \textbf{0.4054} \\
            & 3 & 0.2786 & 0.2100 & 0.1769 & 0.1571 & 0.4380 & 0.2798 & 0.2339 & 0.3881 \\
\bottomrule
\end{tabular}}
\end{table*}
\subsection{Results of Automatic Evaluation}
\subsubsection{Comparison of Baseline and Our Method}
Table \ref{tab:automatic_results} presents the automatic evaluation results of both the baseline LLaVa-1.5 model and our EMCEE model when we prompt them with 0 to 3 example conversations. Our model exhibits substantial improvements over LLaVa-1.5 in terms of both BLEU and ROUGE scores. Specifically, EMCEE shows an increase of 81.6\% in BLEU scores and 80.5\% in ROUGE scores compared to LLaVa-1.5. These results suggest that our model, trained on self-generated synthetic conversations in a multi-round setting, can better explain static XAI and produce responses more aligned with human answers to users' inquiries.

\begin{table*}
\centering
\caption{An ablation study of the proposed EMCEE on the conversational explanation dataset}
\label{tab:ablation_study}
\resizebox{\linewidth}{!}{
\begin{tabular}{ccccccccc}
\toprule
         Methods & BLEU-1 & BLEU-2 & BLEU-3 & BLEU-4 & ROUGE-1 & ROUGE-2 & ROUGE-3 & ROUGE-L \\
\midrule
            
            EMCEE & \textbf{0.3056} & \textbf{0.2336} & \textbf{0.1945} & \textbf{0.1721} & \textbf{0.4629} & \textbf{0.2964} & \textbf{0.2454} & \textbf{0.4054} \\
            No Multi-round Training & 0.2808 & 0.2079 & 0.1685 & 0.1465 & 0.4198 & 0.2608 & 0.2162 & 0.3756  \\
            No Repetition Penalty  & 0.2824 & 0.2214 & 0.1854 & 0.1657 & 0.4219 & 0.2778 & 0.2329 & 0.3798 \\
            No Hallucination Detection & 0.2730 & 0.1977 & 0.1631 & 0.1408 & 0.4161 & 0.2375 & 0.1950 & 0.3625 \\
\bottomrule
\end{tabular}}
\vspace{-10pt}
\end{table*}

\begin{figure}[t]
\centering
    \begin{minipage}[t]{0.8\columnwidth}
        {\centering\includegraphics[width=\linewidth]{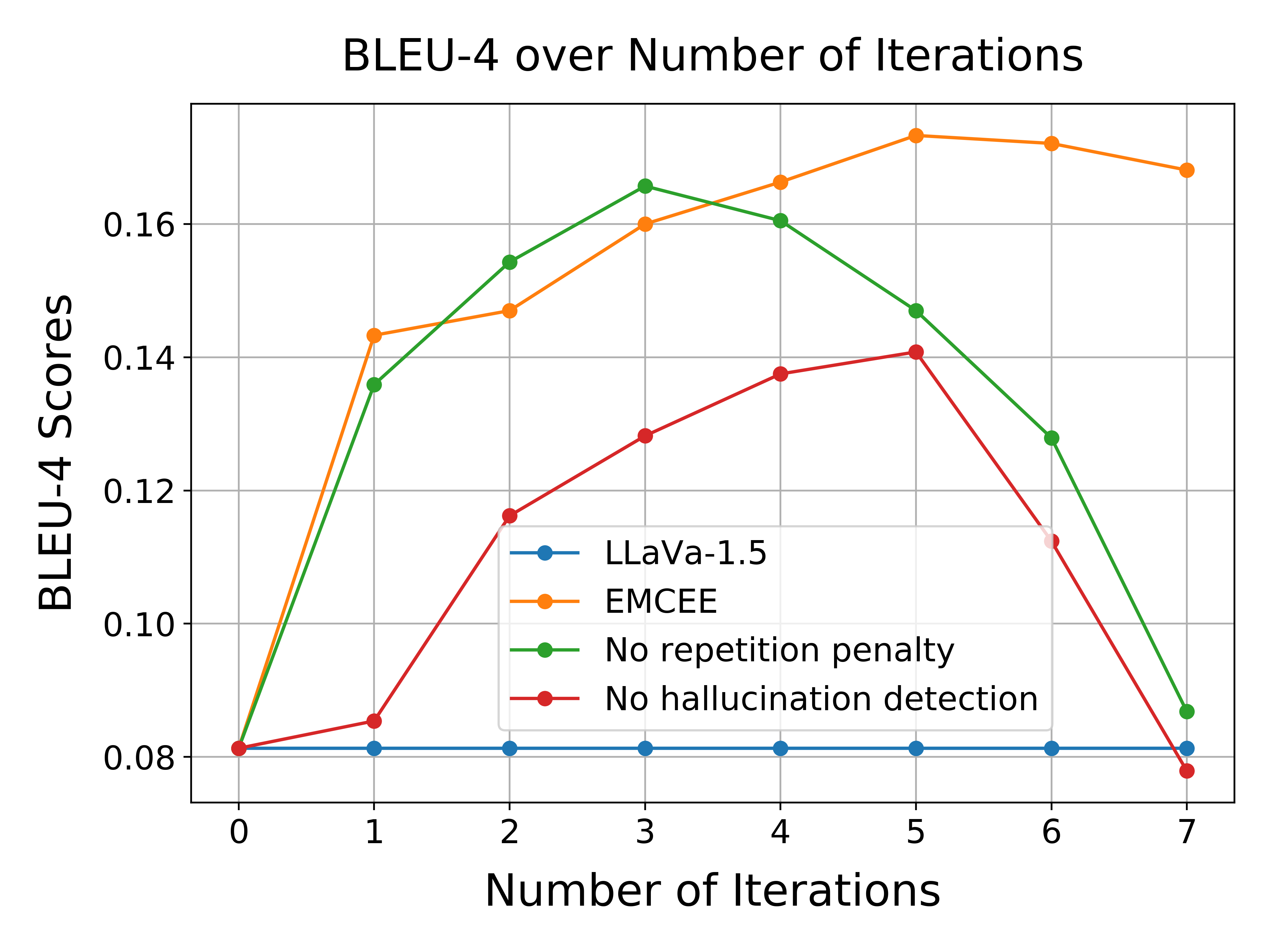}
            \label{bleu_over_iterations}
        }
    \end{minipage}
    \begin{minipage}[t]{0.8\columnwidth}
        {\centering\includegraphics[width=\linewidth]{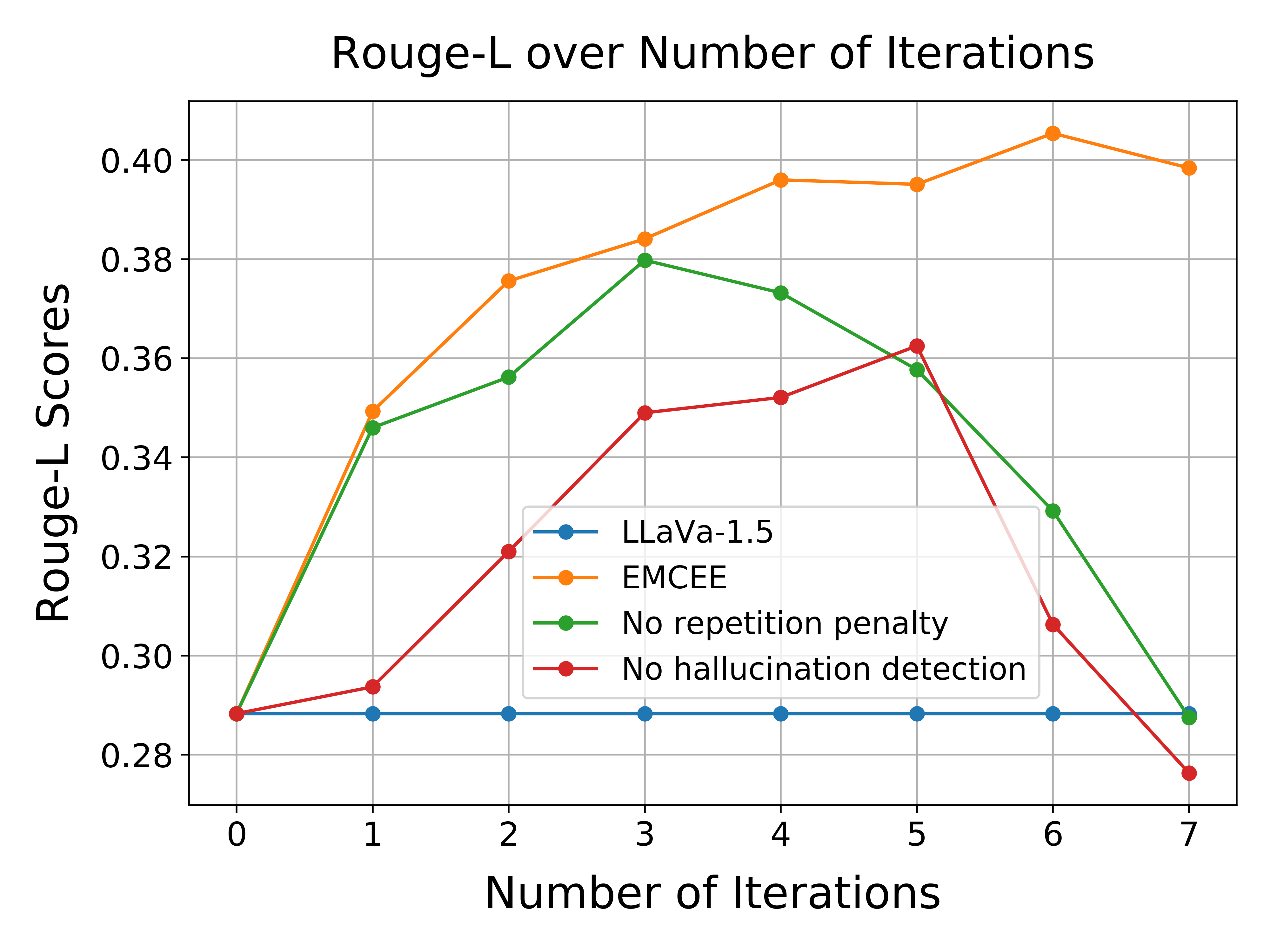}
            \label{rouge_over_iterations}
        }
    \end{minipage}
    \caption{BLEU-4 and Rouge-L scores over the number of training iterations for LLaVa-1.5, EMCEE and different ablated version of EMCEE.}
    \label{fig:score_iterations}
\end{figure}

\subsubsection{Ablation Study}
We created the following ablated versions of EMCEE: (1) No multi-round training, which performs one round of synthetic generation, filtering, and model finetuning. (2) No repetition penalty, which removes the repetition penalty. (3) No hallucination detection, which does not detect and remove hallucinated conversation turns.

Table~\ref{tab:ablation_study} summarizes the results of different ablated versions of EMCEE. We make the following observations. First, the absence of multi-round training significantly reduces the performance across all BLEU and ROUGE metrics. This demonstrates that generating synthetic conversations and filtering out hallucination conversations in an iterative way can gradually improve the quality of generated conversations and thus improve the performance of our model. Second, the model's performance decreases when the repetition penalty is removed. This result indicates that the diversity of synthetic conversations plays a crucial role in our model. Third, the most substantial performance drop occurs when the hallucination detector is removed, with a 10.7\% decrease in BLEU scores and a 15.3\% decrease in ROUGE scores. This result highlights the importance and necessity of filtering hallucinated synthetic data after generation. 

\subsubsection{Effects of Multiple Generation-Training Iterations}
\label{sec:slowed-degeneration}
In the training of EMCEE, we repeated the generation-training process multiple times. We investigated how iterations affect the performance of EMCEE and ablated versions of EMCEE in BLEU-4 and ROUGE-L scores, as shown in Figure \ref{fig:score_iterations}.

We observed that the ablated versions of EMCEE improved in the first few iterations and decreased afterward. This is similar to the findings of \citet{briesch2023large}, who showed that repeatedly training models with self-generated data initially caused performance gains but, after a few iterations, resulted in degenerate synthetic data with low diversity and eventual performance drop. This is especially apparent when we removed the repetition penalty or the hallucination filter, as both BLEU-4 and ROUGE-L decreased drastically after the third and fifth iterations, respectively. However, with both the repetition penalty and the hallucination filter of EMCEE, the performance drops became substantially milder. For BLEU-4, a small drop was observed after the fifth iteration. For Rouge-L, the performance effectively plateaued around the sixth and seventh iteration. We conclude that the proposed techniques, including the repetition penalty and the hallucination filter successfully slow down degeneracy in training with synthetic data.

\begin{table*}
    \caption{Results of human evaluations before and after conversations. Each score is presented as mean $\pm$ standard deviation and the change $\delta = \text{after} - \text{before}$.}
    \centering
    \resizebox{\textwidth}{!}{
    \begin{tabular}{c|c|c|c|c|c|c|c|c}
        \toprule[1pt]
        \multirow{3}*[-0.5em]{\makecell{Explanation\\Methods}} & \multirow{3}*[-0.5em]{\makecell{Conversational\\Explanation\\method}} & \multirow{3}*[-0.5em]{\makecell{Evaluation\\Timing}} & \multirow{3}*{\makecell{Objective\\Understanding\\(Model Selection\\Accuracy)}} & \multirow{3}*[-0.5em]{\makecell{Subjective\\Understanding} } & \multicolumn{3}{c|}{\multirow{2}*{Acceptance}} & \multirow{3}*[-0.5em]{Trust}\\
        &&&&&\multicolumn{2}{c}{}&&\\
        \cline{6-8}
        &&&&&\makecell{Perceived\\Usefulness} & \makecell{Perceived\\Ease of Use} & \makecell{Behavioral\\Intention}&\\
        \midrule
        \multirow{4}*[-1.5em]{LIME} & \multirow{2}*[-0.65em]{Control} 
                            & before & 0.36 $\pm$ 0.09 & 4.60 $\pm$ 1.67 & 5.40 $\pm$ 1.19 & 4.73 $\pm$ 1.04 & 4.90 $\pm$ 0.55 & 4.05 $\pm$ 0.97\\ 
                            & & after& 0.32 $\pm$ 0.18 & 4.40 $\pm$ 1.52 & 5.13 $\pm$ 1.26 & 4.27 $\pm$ 1.30 & 4.60 $\pm$ 0.42 & 3.95 $\pm$ 1.71\\ 
                            & & $\delta$ & -0.04 & -0.20 & -0.27 & -0.46 & -0.30 & -0.10\\
                            \cmidrule(l){2-9}
                            & \multirow{2}*[-0.65em]{LLaVa-1.5} 
                            & before & 0.36 $\pm$ 0.17 & 4.00 $\pm$ 1.58 & 5.20 $\pm$ 1.02 & 4.40 $\pm$ 1.62 & 4.90 $\pm$ 1.02 & 4.10 $\pm$ 0.22\\ 
                            & & after& 0.44 $\pm$ 0.17 & 4.80 $\pm$ 1.48 & 5.60 $\pm$ 0.60 & 5.20 $\pm$ 0.60 & 5.20 $\pm$ 0.76 & 4.30 $\pm$ 0.54\\ 
                            & & $\delta$ & 0.08 & 0.80 & 0.40 & 0.80 & 0.30 & 0.20\\
                            \cmidrule(l){2-9}
                            & \multirow{2}*[-0.65em]{\makecell{EMCEE\\(Ours)}}
                            & before & 0.36 $\pm$ 0.09 & 4.20 $\pm$ 0.84 & 5.27 $\pm$ 0.64 & 4.53 $\pm$ 0.60 & 5.00 $\pm$ 0.35 & 4.20 $\pm$ 0.37\\ 
                            & & after& 0.52 $\pm$ 0.11 & 5.20 $\pm$ 0.45 & 5.93 $\pm$ 0.64 & 5.60 $\pm$ 0.68 & 5.70 $\pm$ 0.45 & 4.85 $\pm$ 0.34\\ 
                            & & $\delta$ & 0.16 & 1.00 & 0.66 & 1.07 & 0.70 & 0.65\\
                            \midrule
        \multirow{4}*[-1.5em]{Grad-CAM} & \multirow{2}*[-0.65em]{Control} 
                            & before & 0.80 $\pm$ 0.14 & 4.00 $\pm$ 1.00 & 5.27 $\pm$ 0.36 & 4.67 $\pm$ 0.67 & 5.00 $\pm$ 0.61 & 4.30 $\pm$ 0.37\\ 
                            & & after& 0.84 $\pm$ 0.17 & 4.00 $\pm$ 1.22 & 5.20 $\pm$ 0.84 & 4.27 $\pm$ 0.92 & 4.90 $\pm$ 0.82 & 4.40 $\pm$ 0.80\\ 
                            & & $\delta$ & 0.04 & 0.00 & -0.07 & -0.40 & -0.10 & 0.10\\
                            \cmidrule(l){2-9}
                            & \multirow{2}*[-0.65em]{LLaVa-1.5} 
                            & before & 0.76 $\pm$ 0.17 & 4.00 $\pm$ 1.41 & 5.33 $\pm$ 0.41 & 4.87 $\pm$ 0.38 & 5.20 $\pm$ 0.57 & 4.40 $\pm$ 0.29\\ 
                            & & after& 0.84 $\pm$ 0.09 & 4.80 $\pm$ 0.45 & 5.60 $\pm$ 0.44 & 5.13 $\pm$ 0.51 & 5.50 $\pm$ 0.50 & 5.00 $\pm$ 0.47\\ 
                            & & $\delta$ & 0.08 & 0.80 & 0.27 & 0.26 & 0.30 & 0.60\\
                            \cmidrule(l){2-9}
                            & \multirow{2}*[-0.65em]{\makecell{EMCEE\\(Ours)}}
                            & before & 0.80 $\pm$ 0.20 & 4.00 $\pm$ 1.22 & 5.13 $\pm$ 1.07 & 4.80 $\pm$ 0.77 & 5.20 $\pm$ 0.27 & 4.15 $\pm$ 0.72\\ 
                            & & after& 0.92 $\pm$ 0.11 & 5.40 $\pm$ 0.89 & 6.13 $\pm$ 0.61 & 5.40 $\pm$ 0.93 & 6.10 $\pm$ 0.42 & 5.25 $\pm$ 0.90\\ 
                            & & $\delta$ & 0.12 & 1.40 & 1.00 & 0.60 & 0.90 & 1.10\\
                            \midrule
        \multirow{4}*[-1.5em]{\makecell{Integrated\\Gradients}} & \multirow{2}*[-0.65em]{Control} 
                            & before & 0.20 $\pm$ 0.20 & 3.80 $\pm$ 0.45 & 4.80 $\pm$ 0.50 & 3.87 $\pm$ 0.90 & 4.20 $\pm$ 0.97 & 3.65 $\pm$ 0.45\\ 
                            & & after& 0.24 $\pm$ 0.17 & 4.00 $\pm$ 0.71 & 4.73 $\pm$ 0.76 & 3.80 $\pm$ 0.77 & 4.00 $\pm$ 1.17 & 3.65 $\pm$ 0.72\\ 
                            & & $\delta$ & 0.04 & 0.20 & -0.07 & -0.07 & -0.20 & 0.00\\
                            \cmidrule(l){2-9}
                            & \multirow{2}*[-0.65em]{LLaVa-1.5} 
                            & before & 0.24 $\pm$ 0.09 & 3.80 $\pm$ 0.45 & 4.73 $\pm$ 0.49 & 3.87 $\pm$ 0.77 & 4.40 $\pm$ 1.08 & 3.85 $\pm$ 0.55\\ 
                            & & after& 0.28 $\pm$ 0.18 & 4.00 $\pm$ 1.00 & 5.00 $\pm$ 0.71 & 4.40 $\pm$ 1.60 & 4.70 $\pm$ 1.20 & 3.85 $\pm$ 0.22\\ 
                            & & $\delta$ & 0.04 & 0.20 & 0.27 & 0.53 & 0.30 & 0.00\\
                            \cmidrule(l){2-9}
                            & \multirow{2}*[-0.65em]{\makecell{EMCEE\\(Ours)}}
                            & before & 0.20 $\pm$ 0.14 & 3.60 $\pm$ 1.14 & 4.67 $\pm$ 0.71 & 3.60 $\pm$ 0.44 & 4.60 $\pm$ 0.65 & 3.85 $\pm$ 0.22\\ 
                            & & after& 0.44 $\pm$ 0.09 & 4.60 $\pm$ 0.55 & 5.20 $\pm$ 0.61 & 4.73 $\pm$ 0.55 & 5.50 $\pm$ 0.71 & 4.50 $\pm$ 0.40\\ 
                            & & $\delta$ & 0.24 & 1.00 & 0.53 & 1.13 & 0.90 & 0.65\\
                            \midrule
        \multirow{4}*[-1.5em]{SHAP}  & \multirow{2}*[-0.65em]{Control}
                            & before & 0.44 $\pm$ 0.17 & 4.20 $\pm$ 0.84 & 5.20 $\pm$ 0.38 & 4.47 $\pm$ 0.56 & 4.80 $\pm$ 0.27 & 4.20 $\pm$ 0.57\\ 
                            & & after& 0.48 $\pm$ 0.23 & 4.00 $\pm$ 1.00 & 5.07 $\pm$ 0.80 & 4.33 $\pm$ 0.85 & 4.70 $\pm$ 0.76 & 4.30 $\pm$ 0.54\\ 
                            & & $\delta$ & 0.04 & -0.20 & -0.13 & -0.14 & -0.10 & 0.10\\
                            \cmidrule(l){2-9}
                            & \multirow{2}*[-0.65em]{LLaVa-1.5} 
                            & before & 0.48 $\pm$ 0.11 & 3.80 $\pm$ 1.79 & 5.40 $\pm$ 0.49 & 4.87 $\pm$ 1.73 & 5.00 $\pm$ 1.06 & 4.20 $\pm$ 1.47\\ 
                            & & after& 0.60 $\pm$ 0.14 & 5.20 $\pm$ 1.64 & 5.60 $\pm$ 0.44 & 5.67 $\pm$ 0.78 & 5.20 $\pm$ 0.91 & 4.60 $\pm$ 1.14\\ 
                            & & $\delta$ & 0.12 & 1.40 & 0.20 & 0.80 & 0.20 & 0.40\\
                            \cmidrule(l){2-9}
                            & \multirow{2}*[-0.65em]{\makecell{EMCEE\\(Ours)}}
                            & before & 0.48 $\pm$ 0.41 & 3.80 $\pm$ 1.30 & 5.40 $\pm$ 0.60 & 4.60 $\pm$ 0.92 & 5.00 $\pm$ 0.79 & 4.20 $\pm$ 0.91\\ 
                            & & after& 0.80 $\pm$ 0.14 & 5.60 $\pm$ 1.14 & 6.13 $\pm$ 0.69 & 6.00 $\pm$ 0.41 & 5.90 $\pm$ 0.89 & 5.30 $\pm$ 0.82\\ 
                            & & $\delta$ & 0.32 & 1.80 & 0.73 & 1.40 & 0.90 & 1.10\\
        \bottomrule[1pt]
    \end{tabular}
}
    \label{tab:human_evaluation_result}
\end{table*}

\subsection{Results of Human Evaluation}
\label{sec:human_evaluation_result}
Table \ref{tab:human_evaluation_result} presents the human evaluation results, comparing LLaVa-1.5, EMCEE, and the no-conversation group across four explanation methods: LIME, Grad-CAM, Integrated Gradients, and SHAP. \hl{The explanation methods (LIME, Grad-CAM, Integrated Gradients, SHAP) and participant groups (control, LLaVa-1.5, EMCEE) are between-subject variables, while time (before vs. after) is a within-subject variable.} We conducted a three-way Analysis of Variance (ANOVA) to analyze the results.

\hl{To ensure the validity of the ANOVA results, we verified its underlying assumptions: independence, normality, and equal variance. For independence, participants were randomly assigned to one of the three groups and each participated in the study only once. For normality, we performed Shapiro-Wilk tests on participants' subjective understanding of different explanation methods. The resulting p-values were 0.155, 0.171, 0.062, and 0.084, indicating that the normality assumption was met. For equal variance, we conducted Bartlett’s test, which yielded a p-value of 0.376, confirming that this assumption was also satisfied.}

\subsubsection{Effects of Different Conversational XAI Systems on Users' Objective Understanding and Subjective Perception of Static Explanations}
\begin{figure}[!ht]
\centering\subfloat[LIME]{\includegraphics[width=0.49\linewidth]{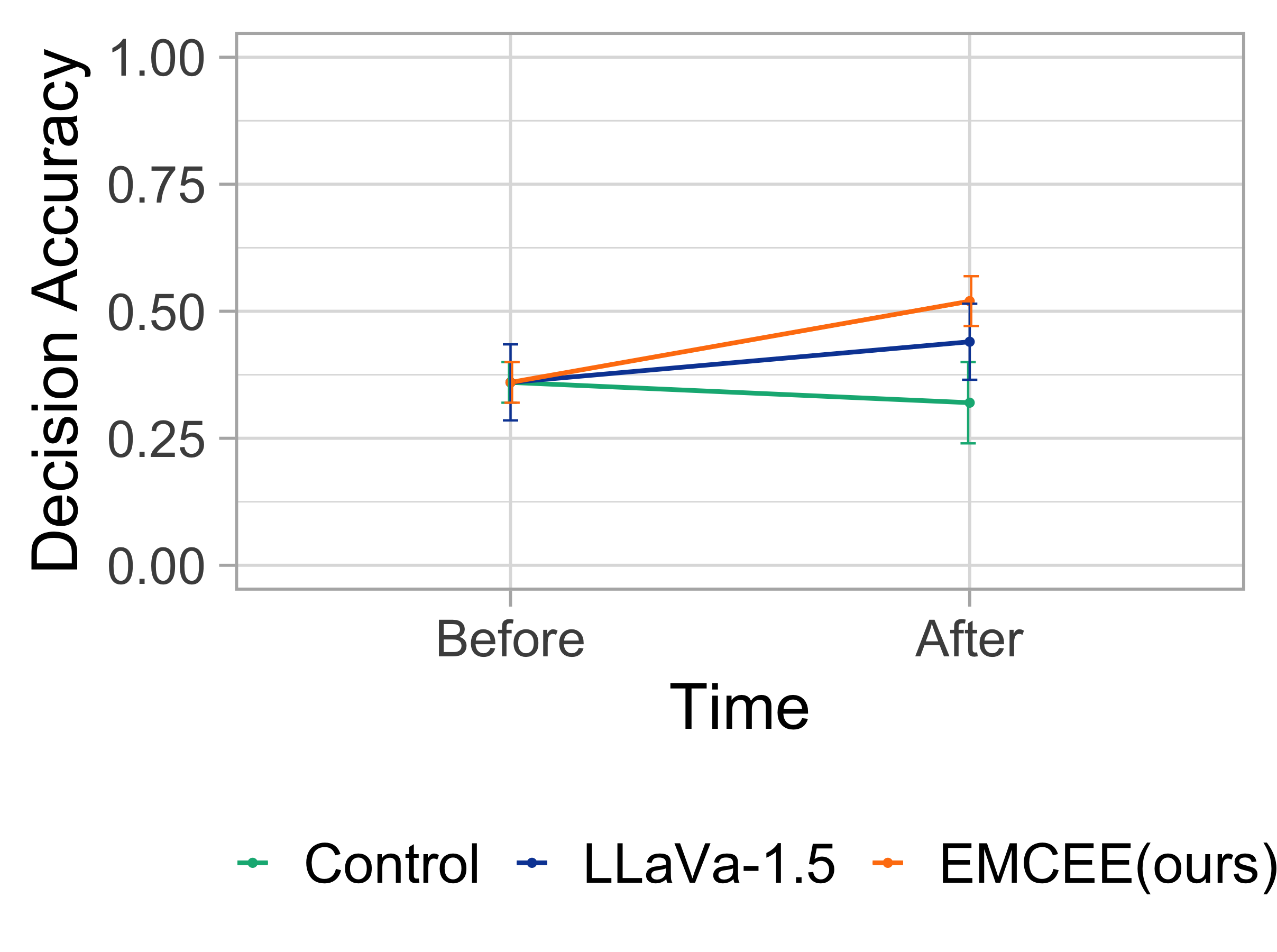}}
\subfloat[Grad-CAM]{\includegraphics[width=0.49\linewidth]{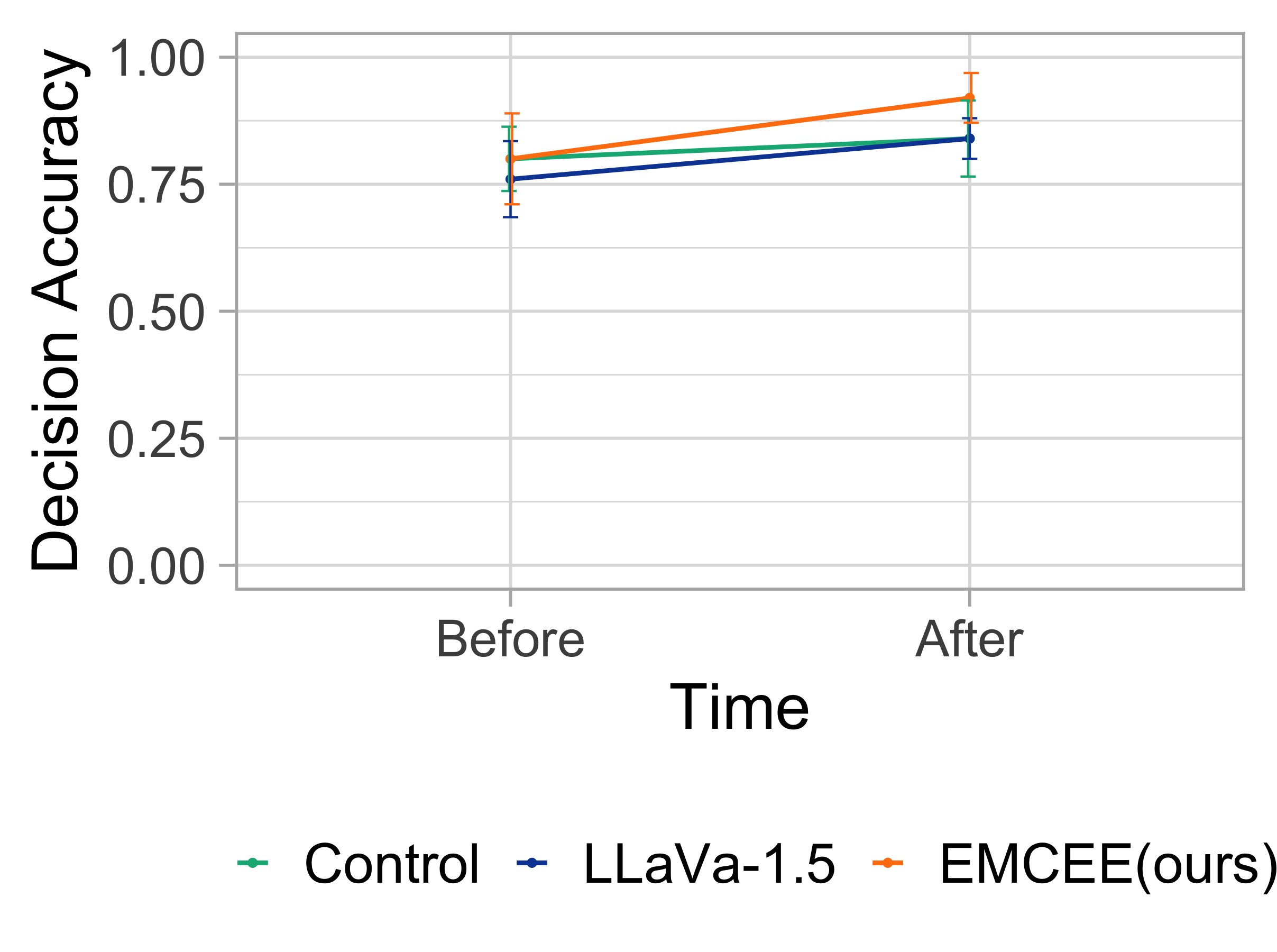}} \hspace{0.04\textwidth}
\subfloat[Integrated Gradients]{\includegraphics[width=0.49\linewidth]{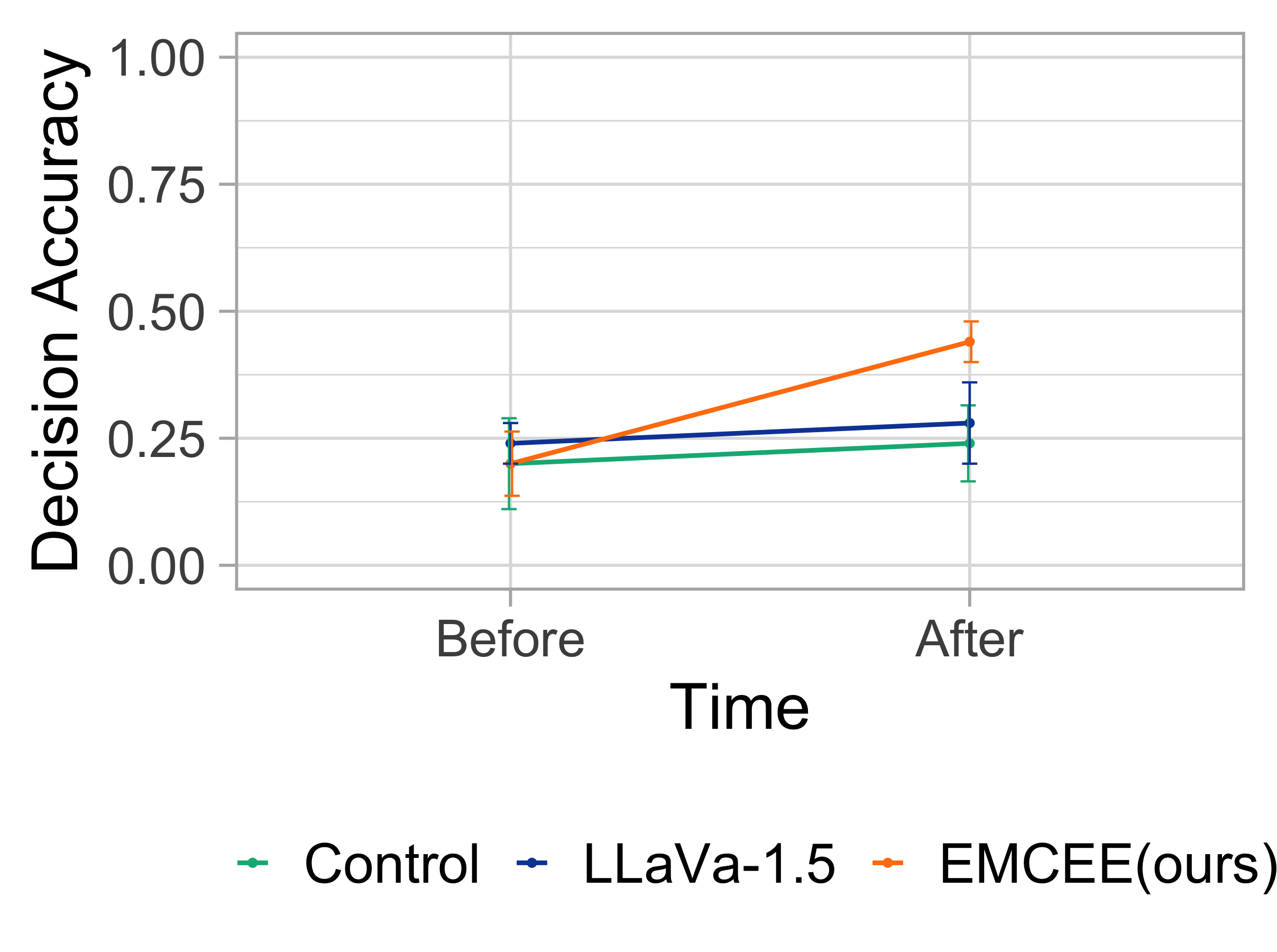}} 
\subfloat[SHAP]{\includegraphics[width=0.49\linewidth]{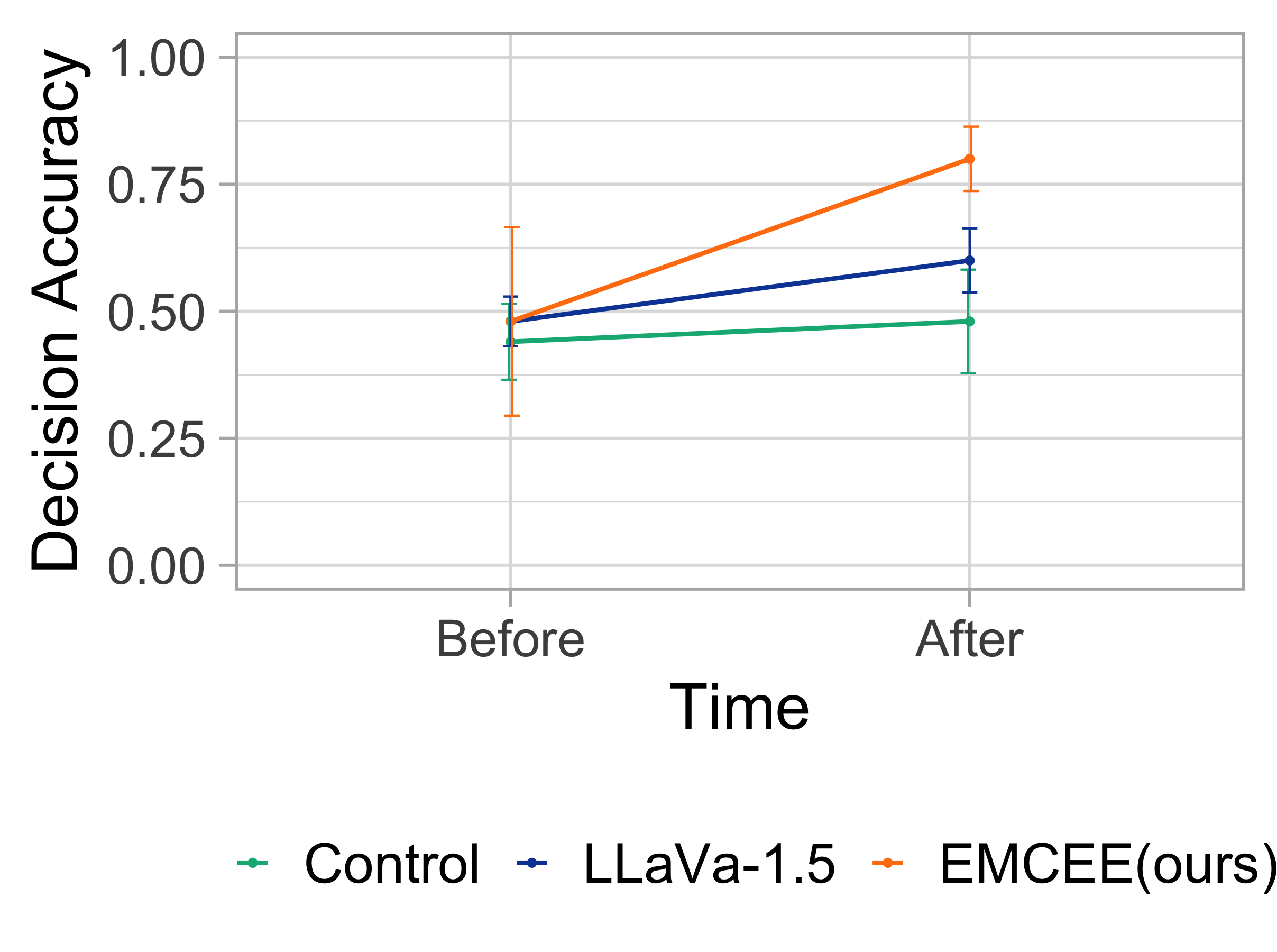}}
\caption{Model selection accuracy for (a) LIME and (b) Grad-CAM (c) Integrated Gradients (d) SHAP before and after conditions. }
\label{fig:interaction_da}
\end{figure}

Results showed significant main effects for group ($F(2,48) = 3.79, p=.04$), method ($F(3,48) = 43.25, p<.001$), and time ($F(1,48) = 18.48, p<.001$). The EMCEE group, the Grad-CAM method, and the after-conversation condition displayed the highest objective decision accuracy. We also found a significant interaction effect between group and time ($F(2,48) = 5.44, p=.007$), as displayed in the Figure \ref{fig:interaction_da}. In participants' initial decisions, no significant differences were observed between the EMCEE, LLaVA-1.5, and control groups. During the final decision, participants interacting with EMCEE or LLaVa-1.5 both showed improved decision accuracy. However, participants using our EMCEE model consistently demonstrated a greater increase in model selection accuracy after the conversation. This phenomenon highlights EMCEE’s effectiveness in helping participants collaborate with static explanations.

We observed varied objective performance across explanation methods ($F(3,48) = 43.25, p<.001$). Participants achieved the highest accuracy in the model selection task with Grad-CAM and the lowest accuracy with Integrated Gradients. A potential reason might be the inherently intuitive nature of the explanations produced by Grad-CAM compared to others \citep{zhang2023may}.

\begin{figure}[!ht]
\centering\subfloat[LIME]{\includegraphics[width=0.49\linewidth]{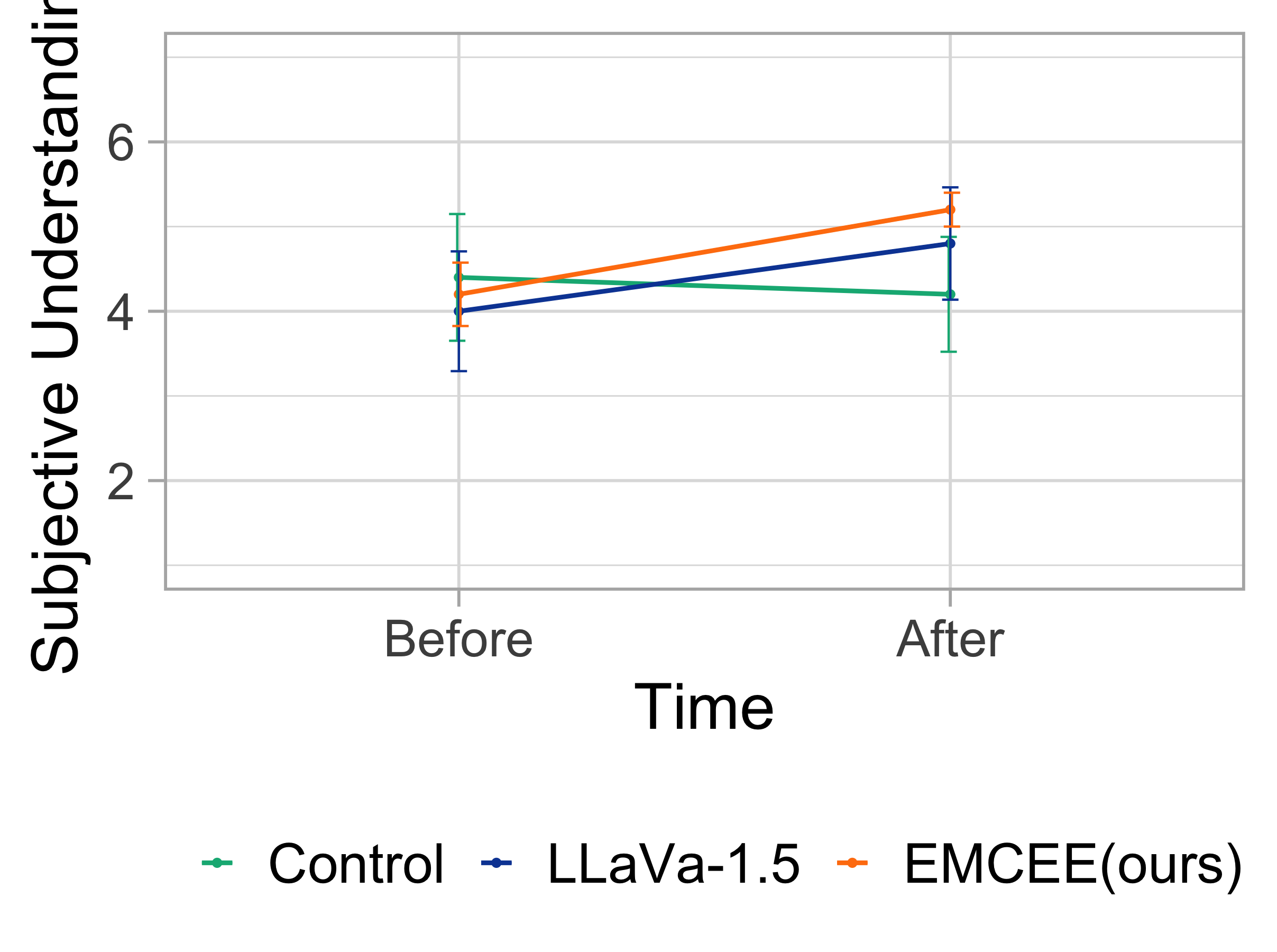}} 
\subfloat[Grad-CAM]{\includegraphics[width=0.49\linewidth]{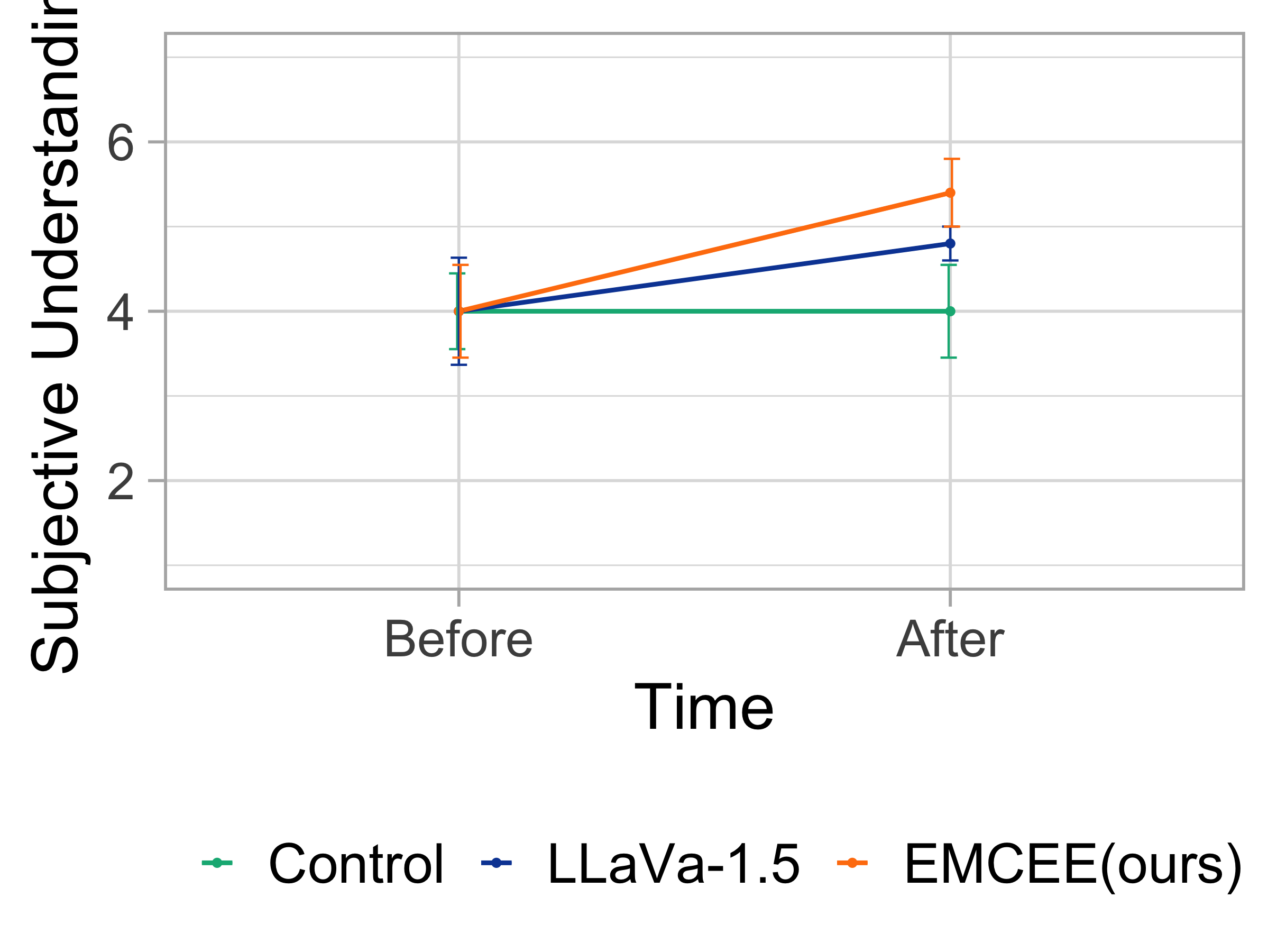}} \hspace{0.04\textwidth}
\subfloat[Integrated Gradients]{\includegraphics[width=0.49\linewidth]{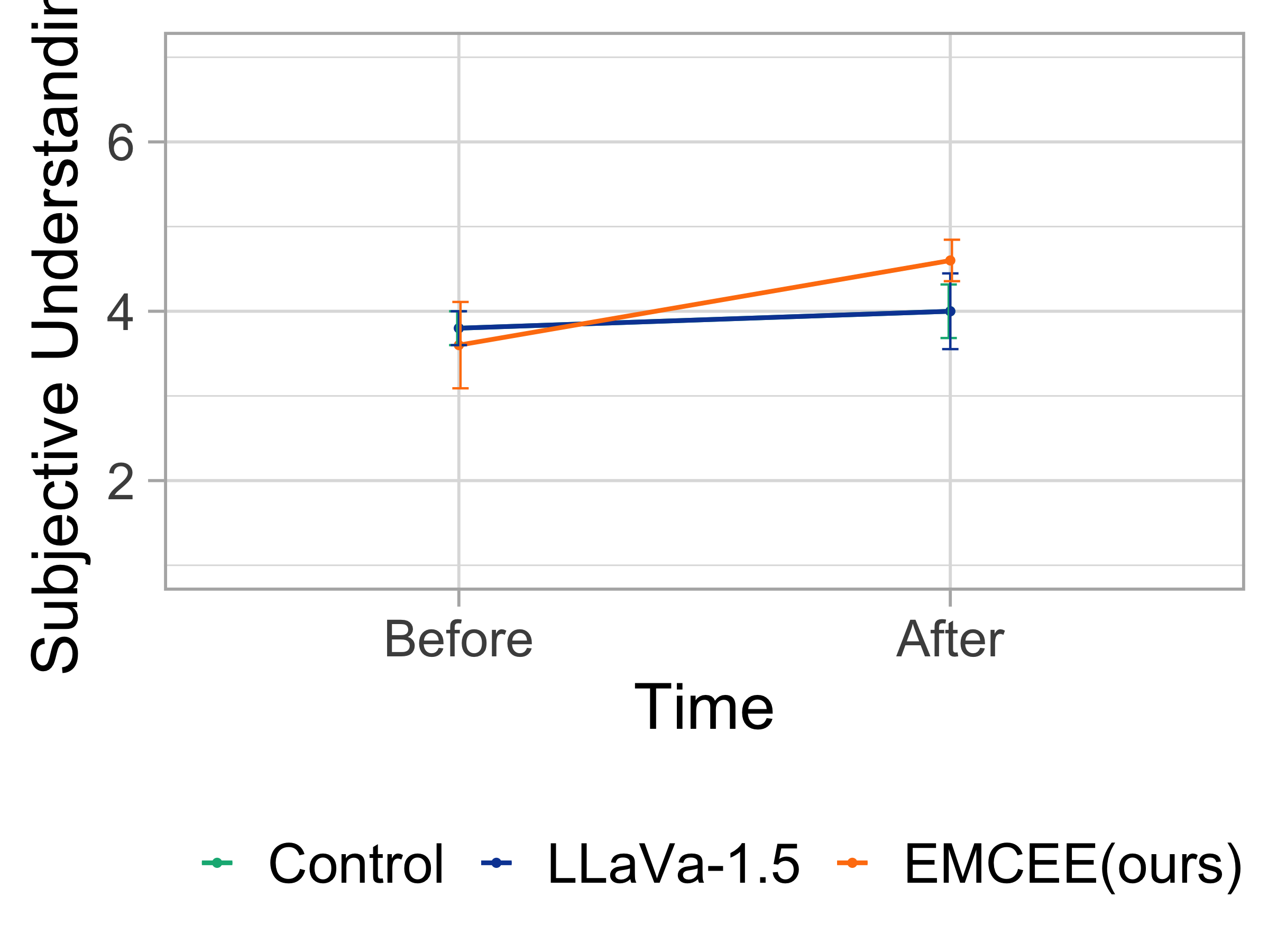}} 
\subfloat[SHAP]{\includegraphics[width=0.49\linewidth]{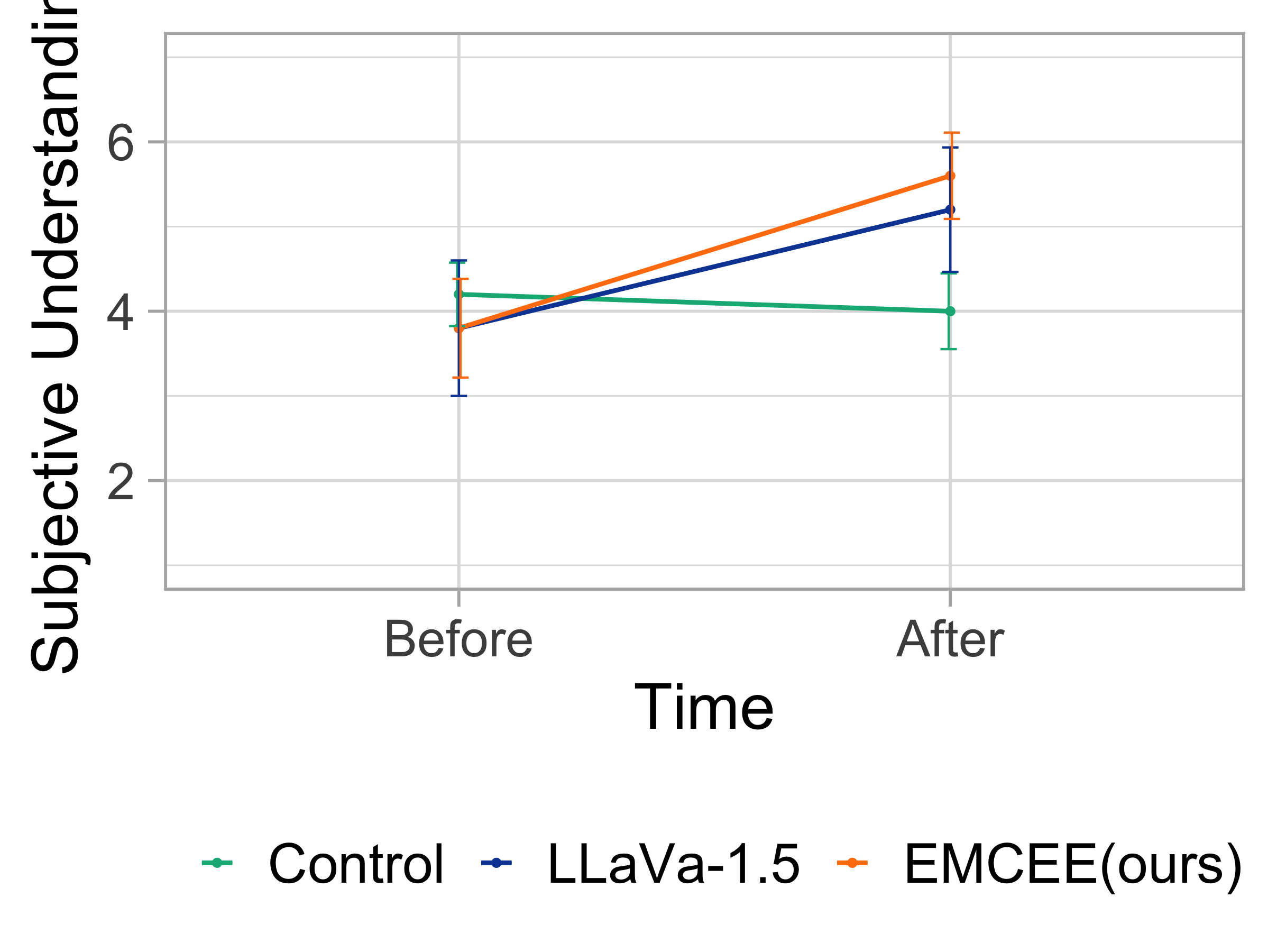}}
\caption{Subjective understanding score for (a) LIME and (b) Grad-CAM (c) Integrated Gradients (d) SHAP before and after conditions. }
\label{fig:su_interaction}
\end{figure}

Regarding participants' subjective understanding, we found a significant main effect of evaluation timing ($F(1,48) = 30.56, p< .001$) and a significant interaction between group and time ($F(1,116) = 10.16, p < .001$). Initially, there was no significant difference in participants’ self-reported understanding of static explanations among different groups. After the conditions, participants who received conversational explanations from EMCEE reported significantly greater improvements than the other two groups across all four explanation methods.

\begin{figure}[!ht]
\centering\subfloat[LIME]{\includegraphics[width=0.49\linewidth]{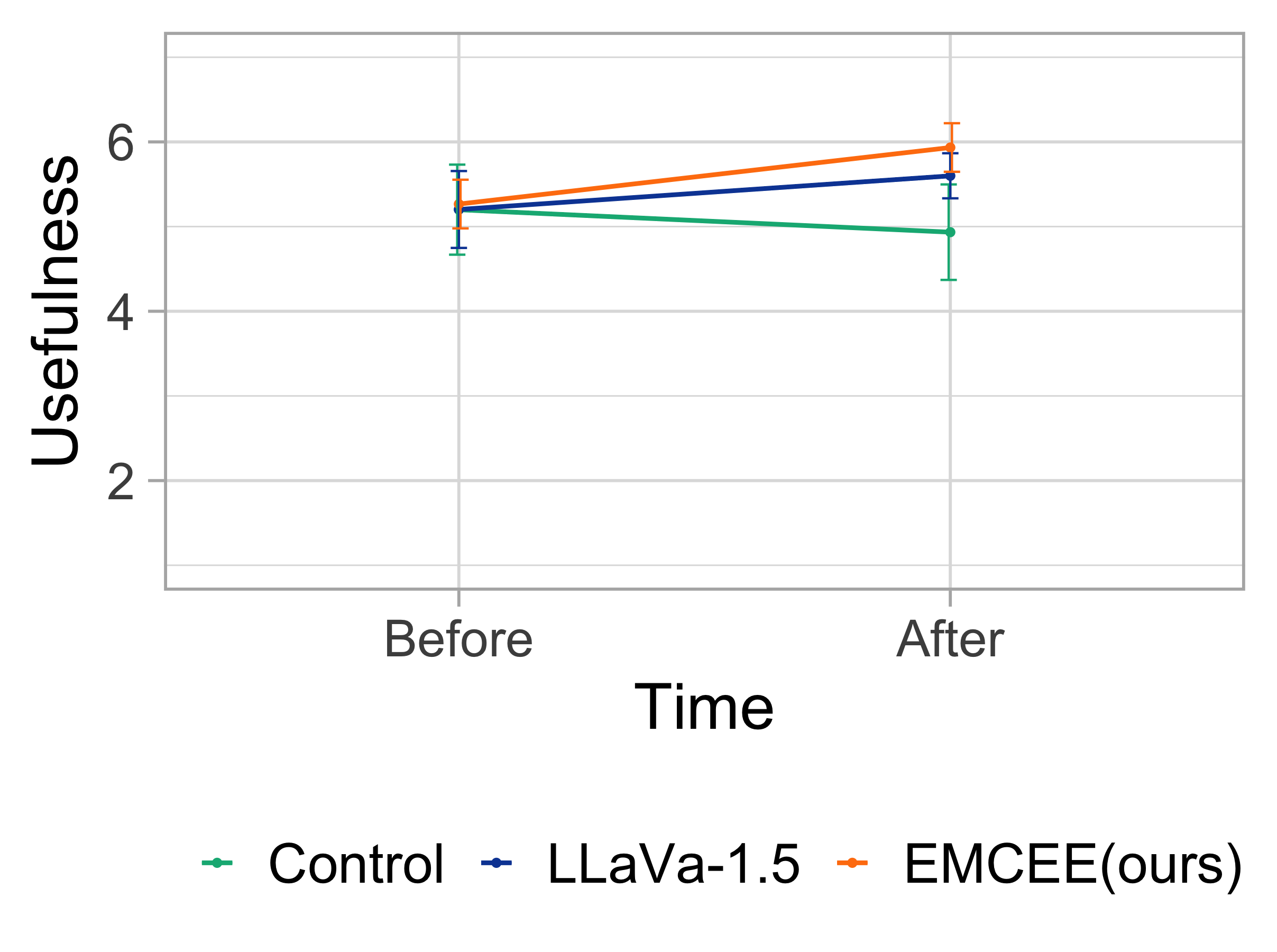}} 
\subfloat[Grad-CAM]{\includegraphics[width=0.49\linewidth]{img/U_LIME.png}} \hspace{0.04\textwidth}
\subfloat[Integrated Gradients]{\includegraphics[width=0.49\linewidth]{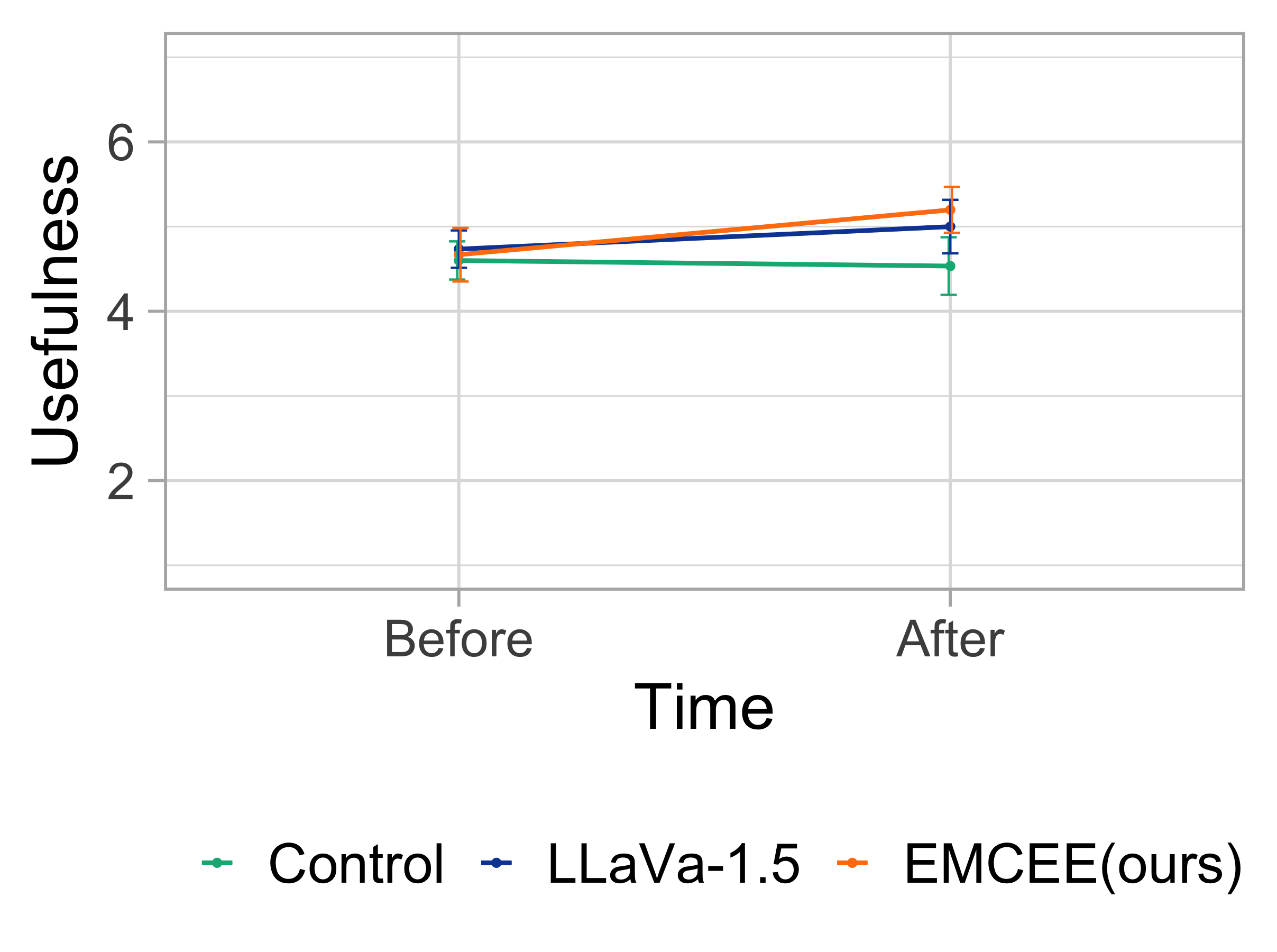}} 
\subfloat[SHAP]{\includegraphics[width=0.49\linewidth]{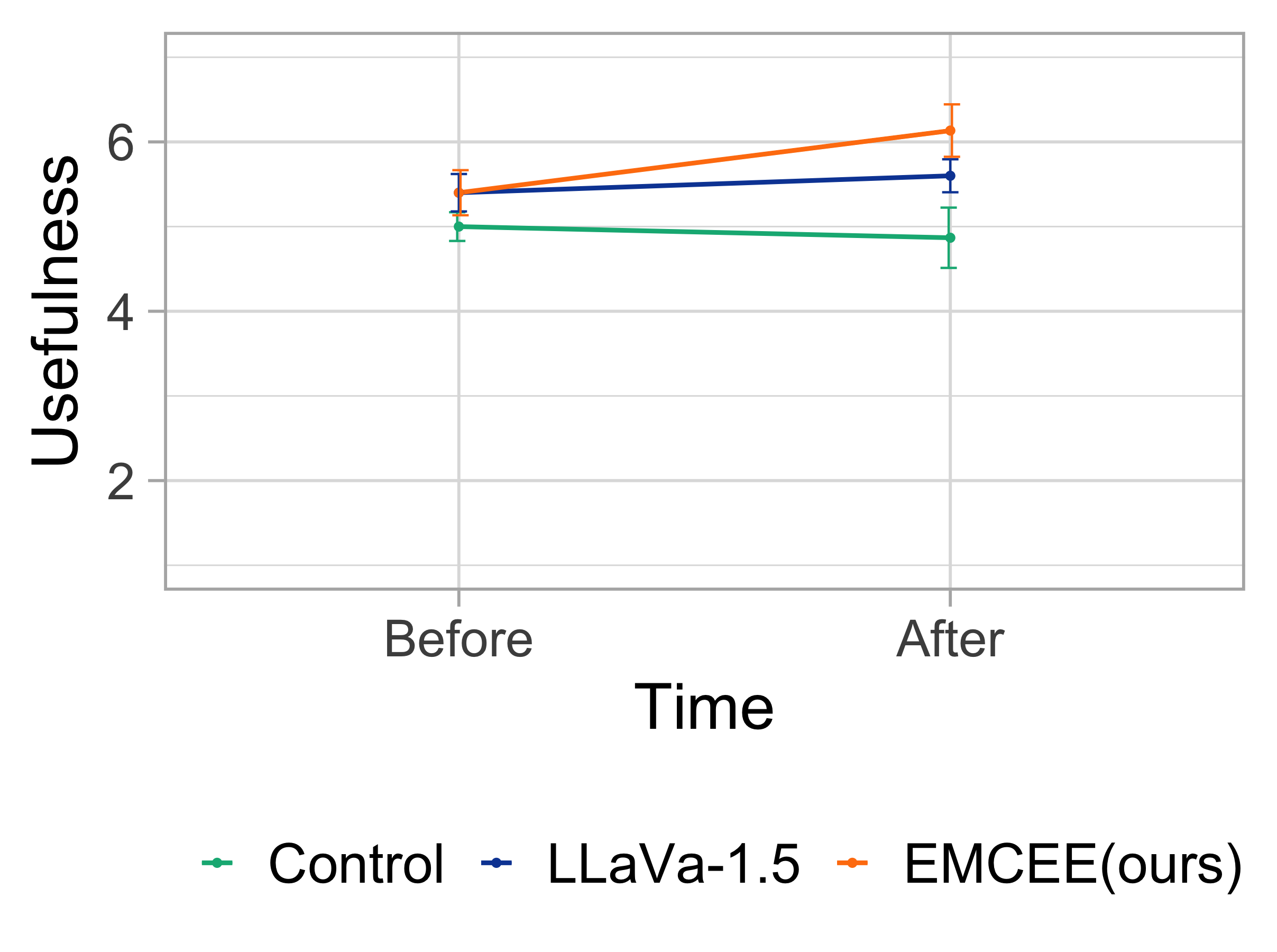}}
\caption{Participants' self-report usefulness score for (a) LIME and (b) Grad-CAM (c) Integrated Gradients (d) SHAP before and after conditions. }
\label{fig:u_interaction}
\end{figure}

For perceived usefulness, the results showed a significant main effect of method ($F(3,48) = 2.86, p = .0046$) and time ($F(1,48) = 21.35, p < .001$), as well as a significant interaction between group and time ($F(2,48) = 15.37, p < .001$), as depicted in Figure \ref{fig:u_interaction}. Participants’ perceived usefulness increased after interacting with LLaVa-1.5 or EMCEE, though the improvement is much smaller with LLaVa-1.5. In contrast, for the control group, perceived usefulness dropped after more time to the static explanations was provided.

Similar results were observed for participants' perceived ease of use. There were significant main effects of method ($F(3,48) = 3.83, p=.002$) and of time ($F(1,48) = 22.14, p<.001$), as well as a significant interaction effect between group and time ($F(2,48) = 15.5, p<.001$). The interaction effect is displayed in appendix Figure \ref{fig:eoe_interaction}. The perceived ease of use increased after participants interacted with EMCEE or LLaVa-1.5. EMCEE produced a greater improvement than LLaVa-1.5. On the contrary, the control group’s perceived ease of use decreased after spending more time with static explanations.

For the behavioral intention, results showed significant main effects of the group ($F(2,48) = 5.14, p=.009$), method ($F(3,48) = 2.84, p=.004$), and time ($F(1,48) = 18.48, p<.001$). We also observed a significant interaction effect between group and time ($F(1,116) = 20.94, p<.001$). The interaction figure is displayed in appendix Figure \ref{fig:bi_interaction}.  Participants are more inclined to use explanations in future scenarios after receiving conversational explanations from EMCEE.
On the other hand, the behavioral intention of the control group decreased for all explanation methods.

The boost in perceived usefulness, ease of use, and behavioral intention after interacting with EMCEE can be attributed to the increased understanding of static explanations. Prior to the interactions, participants might have had limited knowledge or even misconceptions about the explanation methods \cite{zhang2023may}. Experiment results showed that participants gained a clearer understanding of how the XAI methods function, after the participants’ questions were addressed in the conversations with EMCEE. Consequently, they reported perceiving the static explanations as more useful and easier to use, and were more inclined to use the static explanations in future tasks.

\begin{figure}
\centering\subfloat[LIME]{\includegraphics[width=0.49\linewidth]{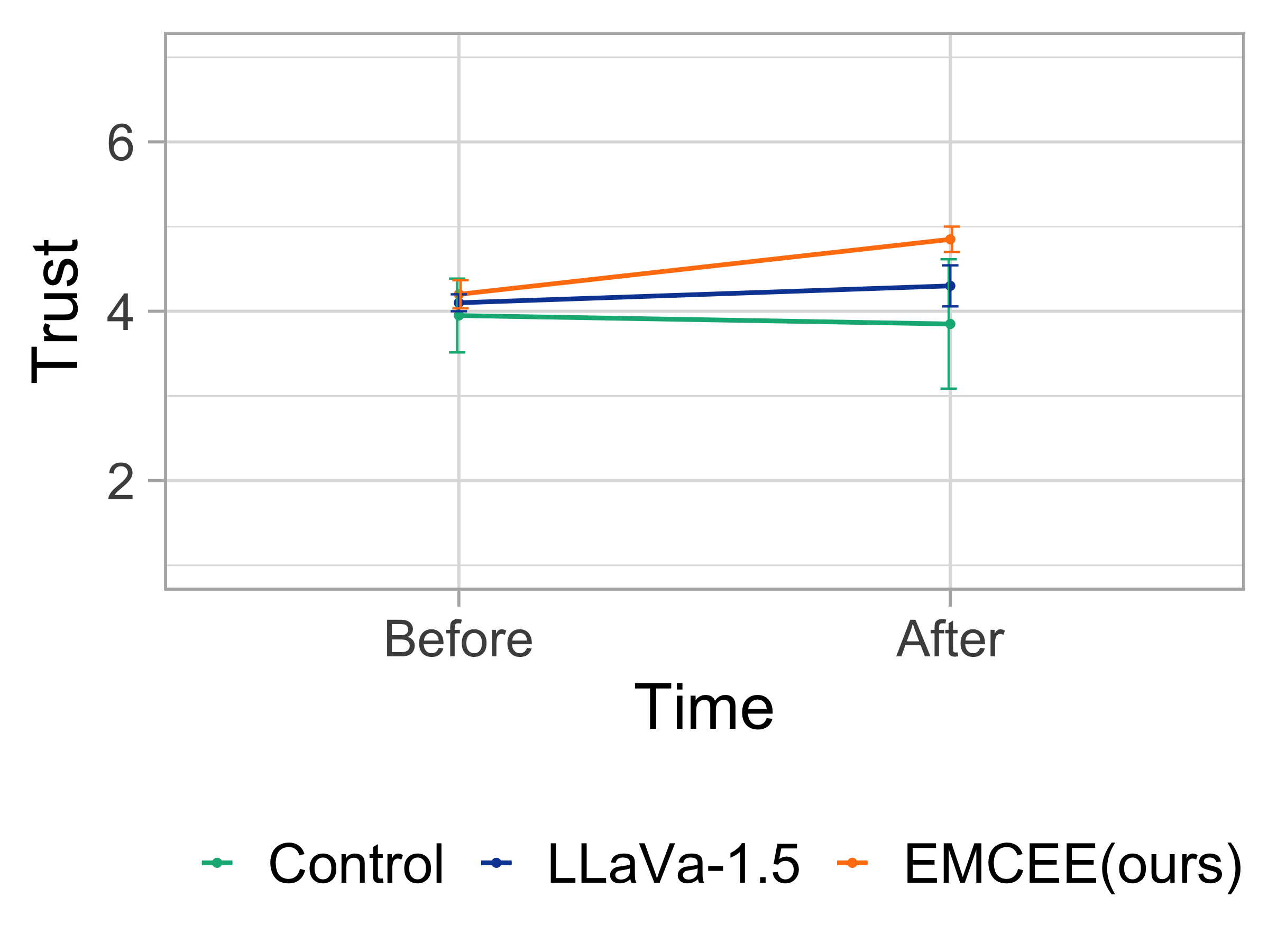}} 
\subfloat[Grad-CAM]{\includegraphics[width=0.49\linewidth]{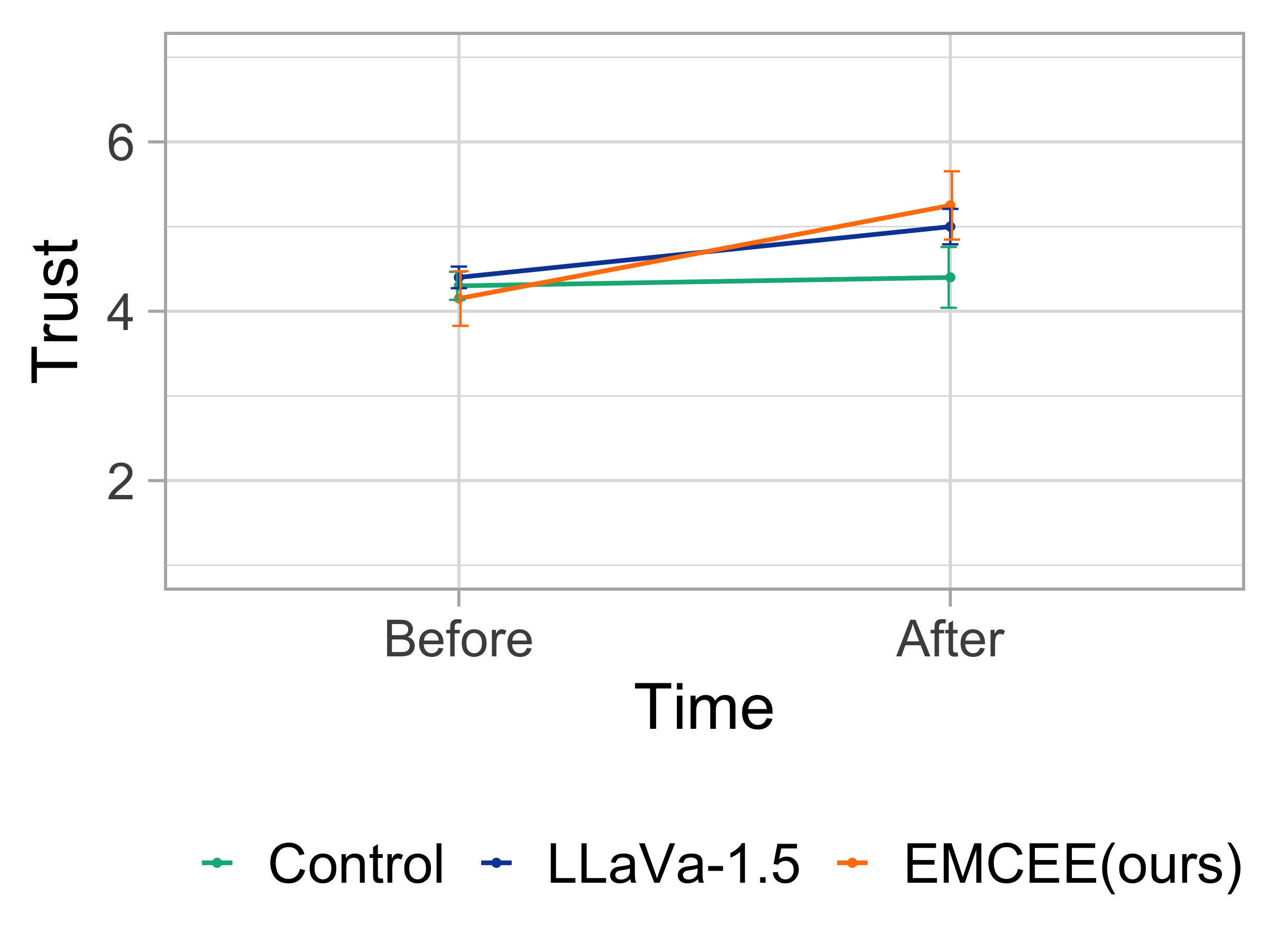}} \hspace{0.04\textwidth}
\subfloat[Integrated Gradients]{\includegraphics[width=0.49\linewidth]{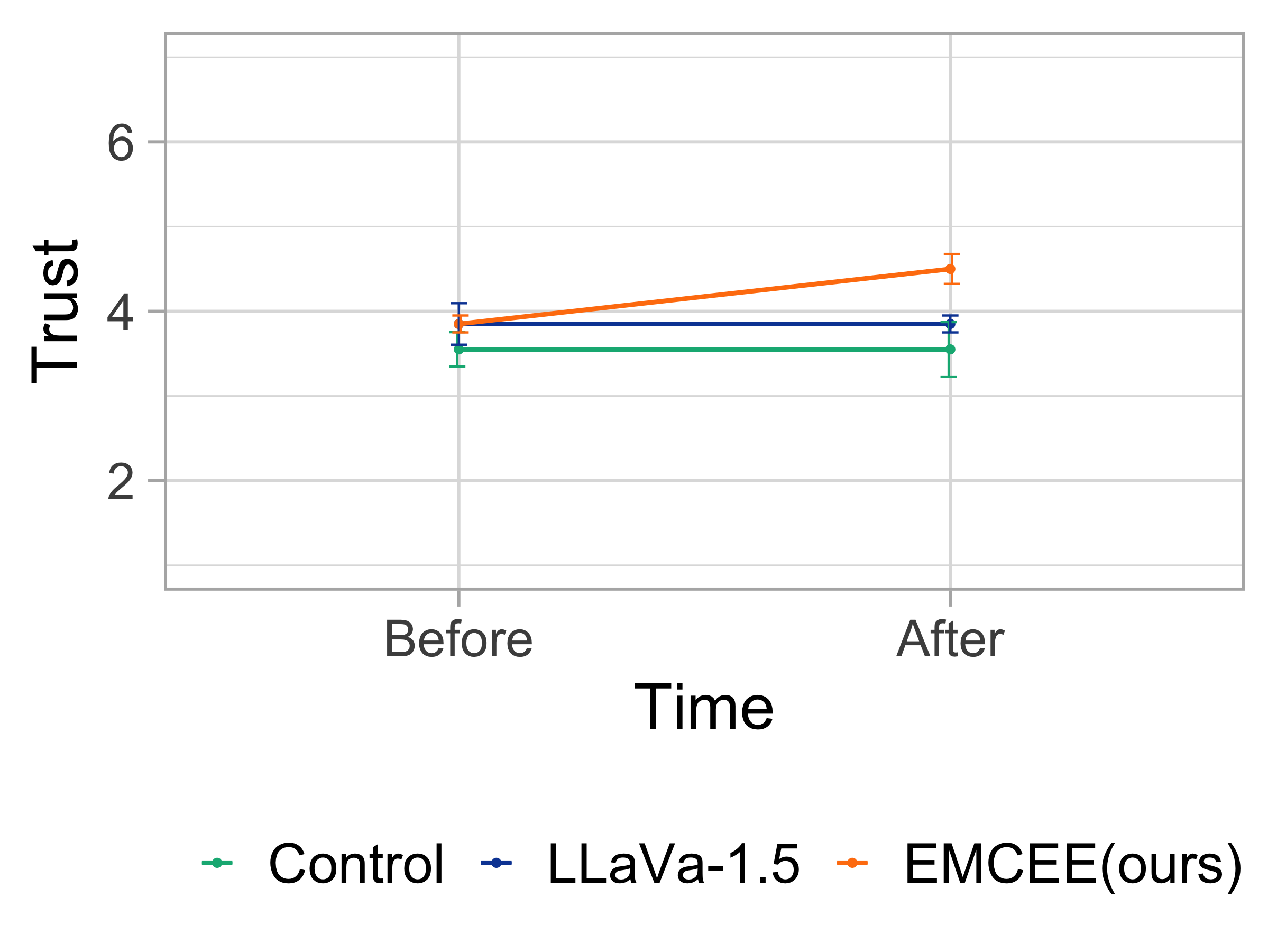}} 
\subfloat[SHAP]{\includegraphics[width=0.49\linewidth]{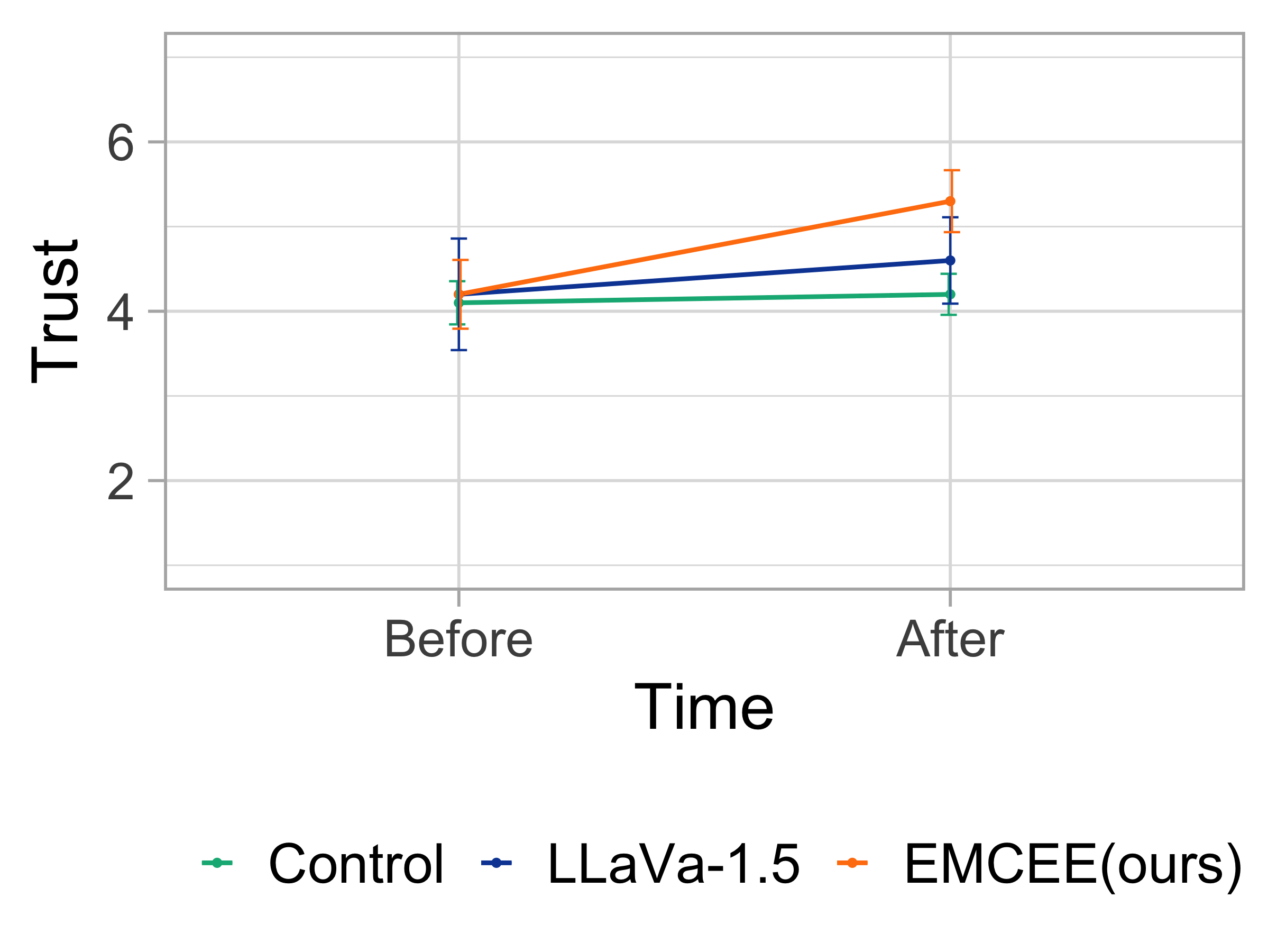}}
\caption{Participants' trust for (a) LIME and (b) Grad-CAM (c) Integrated Gradients (d) SHAP before and after conditions.}
\label{fig:t_interaction}
\end{figure}

For the trust measurement, results showed a significant main effect of time ($F(1,48) = 40.16, p < .001$) and a significant interaction effect between group and time ($F(1,116) = 43.7, p < .001$), as shown in Figure \ref{fig:t_interaction}. Initially, there were no significant differences in trust scores among the groups. However, by the end, participants who interacted with EMCEE reported the highest trust scores. According to theories of trust \citep{mcknight1998initial, lim2009why, hoffman2018metrics}, the ability to build a mental model of AI systems is the key to user trust in AI. The improvements in trust may be a result of an improved understanding of static explanations, as indicated by earlier results. 

\hl{From Table~{\ref{tab:human_evaluation_result}}, one may notice that the improvement deltas are small relative to the standard deviation of the measurements. The relatively large standard deviation is due to inherent variations in individuals' subjective perceptions. These variations arise from differences in participants' backgrounds, experiences, and understanding of explanation methods. Despite this, the deltas capture the overall shift of the entire user group before and after the study, indicating the impact of different experimental conditions. Our deltas are consistently higher than those of the baselines across all explanation methods and measurements, demonstrating the effectiveness of the proposed EMCEE.}

\subsubsection{Analysis of Collected Conversations}
\label{section:analysis_conversation}

\begin{table*}
    \caption{Overview of Collected Questions. Including categories of questions, examples, and the number of questions in each category.}
    \centering
    \resizebox{\linewidth}{!}{
    \begin{tabular}{>{\centering\arraybackslash}m{4cm}|m{12cm}|m{0.8cm}}
    \toprule
        \multicolumn{1}{>{\centering\arraybackslash}m{4cm}|}{Question Category} &\multicolumn{1}{>{\centering\arraybackslash}m{12cm}|}{Question Examples} & \multicolumn{1}{>{\centering\arraybackslash}m{0.8cm}}{Num}\\
        \midrule
        Generic questions about machine learning and explainable AI concepts & 
        \begin{itemize}[leftmargin=2ex, topsep=3pt, parsep=0.2ex, after=\vspace{1pt}, itemsep=-0.25ex,label=\raisebox{0.25ex}{\tiny$\bullet$}]
            \item What is a deep learning model?
            \item What is Swin Transformer?
            \item What is an explanation method?
        \end{itemize}
         & 87 \\
        \midrule
       Questions related to the provided explanations: & 
        \begin{itemize}[leftmargin=2ex, topsep=3pt, parsep=0.2ex, after=\vspace{1pt}, itemsep=-0.25ex,label=\raisebox{0.25ex}{\tiny$\bullet$}]
            \item How does SHAP determine the regions of the image that are most important for the prediction?
            \item How does it mean by the output changes when the input changes (in Integrated Gradients)?
            \item Would the Grad-CAM get wrong?
        \end{itemize}
        & 168 \\
        \midrule
        Extended questions & 
        \begin{itemize}[leftmargin=2ex, topsep=3pt, parsep=0.2ex, after=\vspace{1pt}, itemsep=-0.25ex,label=\raisebox{0.25ex}{\tiny$\bullet$}]
            \item Can I use grad-cam for an image containing more than 1 type of animals?
            \item What if some important or unique parts of the animal are blocked? How should this image be classified, and can you provide such explanations?
            \item What are the potential limitations when using SHAP in practical applications?
        \end{itemize} & 103\\
    \bottomrule
    \end{tabular}}
    \label{tab:questions}
\end{table*}

We collected 40 conversations between participants from four different discipline groups and two conversational explanation systems. On average, each conversation has 22.8 turns. By conducting a basic content analysis of the users' questions, we divide them into three categories:
\begin{itemize}
    \item Generic questions about machine learning and explainable AI concepts: Questions about fundamental terms and concepts in machine learning and explainable AI that lay people may not know. Examples include, "What is a deep learning model?",  "What is accuracy?", or "What are explanation methods?"
    \item Questions related to the provided explanations: Questions about the specific explanations provided during the conversation, such as how the explanation is created and how the explanation methods function. Examples include, "How does Grad-CAM produce the heatmap?" or "What do different colors represent in SHAP?".
    \item Extended questions: Questions that arise after users understand the provided explanations, e.g., generating other explanations for the current prediction, explanations for different predictions, or comparisons between the provided explanation and other explanation methods.
\end{itemize}

Based on this categorization, we classified all questions in our collected conversations. In total, we identified 358 questions across the three categories. Table \ref{tab:questions} provides examples and the number of questions in each category. As observed in Table~\ref{tab:questions}, a large portion of the questions revolve around basic machine learning and explainable AI concepts. This trend might be attributed to the diverse backgrounds of the participants. It suggests that many participants may not be familiar with machine learning models and explanation methods. This is consistent with the real application of explanation methods, where non-expert users often need clarification on fundamental concepts.

We also observed a significant interest of participants in questions related to the provided explanations. This suggests that explanations generated by Grad-CAM, LIME, and Integrated Gradients are not always easily understood by users. This highlights the importance of tailoring responses to users' specific questions to enhance their understanding of these explanations. Furthermore, participants demonstrated notable curiosity regarding extended questions, such as asking for new explanations or comparisons between different explanations. This indicates that as participants become more familiar with the provided explanations, they develop an interest in exploring alternative methods and understanding how models might behave in different scenarios.

\begin{table*}
    \centering
    \resizebox{\textwidth}{!}{
    \begin{tabular}{c|c|c|c|c|c|c}
    \hline
        \multirow{2}*{Model} & \multicolumn{2}{c|}{Question Category 1} & \multicolumn{2}{c|}{Question Category 2} & \multicolumn{2}{c}{Question Category 3}\\
    \cline{2-7}
        & Understandability & \makecell{Factual\\correctness} & Understandability & \makecell{Factual\\correctness} & Understandability & \makecell{Factual\\correctness}\\
    \hline
        LLaVa-1.5   & 0.77 & 0.70 & 0.63 & 0.77 & 0.78 & 0.70\\
        EMCEE(ours) & 0.78 & 0.83 & 0.83 & 0.87 & 0.80 & 0.80\\
    \hline
    \end{tabular}
    }
    \caption{Understandability and Factual Correctness of replies generated by EMCEE and LLaVa-1.5. Two scores are rated as 0 or $1$. The best results are in \textbf{boldface}. We measure the inter-rater reliability with Fleiss’ Kappa \cite{kappa}. Our annotations obtain “moderate agreement” for Understandability (0.57) and “substantial agreement” for Factual Correctness  (0.675).
    }
    \label{tab:label_question_answers}
\end{table*}

To better understand the advantages of the proposed EMCEE model compared to the baseline LLaVa-1.5, we randomly selected 60 question-answer pairs from the conversations collected in the human evaluation. For each model, we selected 10 question-answer pairs from each of the three question categories. We then recruited three well-educated annotators to evaluate the answers based on two criteria: \textit{Factual Correctness} and \textit{Understandability}. Factual correctness assesses whether the responses are accurate, while understandability measures whether the responses are easy to comprehend. Each criterion is rated as either $0$ or $1$. Table \ref{tab:label_question_answers} presents the results, which showed that the EMCEE model consistently generated more factually correct answers across all three categories, compared to the baseline model. This improvement can be attributed to the use of a hallucination detector during the training phase, which removes factually incorrect statements from the synthetic data and reduces the hallucinations in the final model. Regarding understandability, EMCEE outperforms the baseline, particularly in questions related to the provided explanations. This is likely due to the method used for generating synthetic conversations, where both questions and answers are conditioned on the explanations. As a result, when trained on this data, the EMCEE model becomes more adept at answering questions about the explanations provided.

\begin{table}[ht]
    \begin{tabular}
    {p{0.8cm} p{6.8cm}} 
    \hline
    \textbf{Type} & \textbf{Utterance} \\
    \hline
    Q: & Does the XAI provide explanations about how the classification model made the prediction? \\
    A: & No, it only gives local explanations about what the model predicts. \\
    \hline
    Q: & Are there any limitations to the Grad-CAM? \\
    A: & No, there are no limitations to the method. \\
    \hline
    Q: & Is the process of SHAP complicated? \\
    A: & No, SHAP is a relatively simple process, even though it provides deeper feature attribution insights than traditional methods such as permutation importance tests. \\
    \hline
    Q: & Does LIME produce probabilities or classifications? \\
    A: & LIME produces classifications, not probabilities. \\
    \hline
    \end{tabular}
    \caption{Examples of conversation turns that are identities as hallucinations by the detector.}
    \label{tab:hallucination_example}
\end{table}
\subsubsection{Examples of Identified Hallucinations in Generated Conversations}
To qualitatively assess the effectiveness of the hallucination detector, we selected four conversation turns identified as hallucinations by the detector, as presented in Table \ref{tab:hallucination_example}. These examples demonstrate that LLMs tend to generate untruthful responses about both fundamental machine learning concepts and various XAI techniques. The hallucination detector in our model can successfully identify and exclude such incorrect turns from the synthetic dataset. Consequently, the hallucination detection and filtering process diminishes the occurrence of hallucinations in the synthetic data and enhances the performance of models finetuned on this refined dataset.

\section{Limitations}
We identified five limitations of the current work. First, the static explanations used in our study are limited. Our experiments focused on feature attribution explanation methods on image classification. 
Even though our method is applicable to any static explanation method, the performance of our model on other types of static explanation methods, such as example-based explanation methods, or NLP tasks, is yet to be explored. 
\hl{Second, with 79.5\% accuracy on held-out test data, the hallucination detector is not perfect. Errors from the detector may cause hallucinated responses to slip through or valid responses to be incorrectly filtered.}
Third, we mainly focused on removing factuality hallucinations, while not considering faithfulness hallucinations~\citep{huang2023survey}. Factuality hallucinations refer to statements that are factually incorrect or fabricated. Faithfulness hallucinations refer to statements that are not related to instructions and contextual information. In data generation, our model also may generate unrelated conversations to the static explanations. We leave building a detector or using other methods to filter these unrelated conversations for future work. 
\hl{Fourth, previous work {\citep{ehsan2021explainable}} indicates that prior knowledge of AI may influence participants’ perception of explanations. We mitigated this potential confounding factor by randomly assigning participants. However, the effect of prior knowledge on the use of conversational XAI remains an open problem.}
Finally, our research is confined to one geographical region. Factors such as cultural backgrounds could potentially affect how users interact with XAI and how they seek to clarify confusion. Future studies could involve recruiting participants from diverse countries and regions.

\section{Conclusion}
This paper proposes the fEw-shot Multi-round ConvErsational Explanation (EMCEE) to provide customized explanations to users from diverse domains. To deal with data scarcity, we train the EMCEE with synthetic data. We first use a vision language model to generate synthetic conversations with the repetition penalty to promote the diversity of generated data. Then, to reduce hallucinations in generated data, we apply a hallucination detector to filter hallucinated conversation turns after the data generation. To iteratively improve the performance, we repeat the generation-filtering-finetuning process multiple times. Both automatic and human evaluation demonstrate that EMCEE outperforms baseline models by a large margin. In practice, EMCEE significantly improved users' comprehension, acceptance, trust, and collaboration with static explanations. By analyzing conversations, we demonstrate that EMCEE can generate more truthful and understandable responses, leading to a better user experience. 

\section{Acknowledgment}
We gratefully acknowledge the support by the Nanyang Associate Professorship and the National Research Foundation Fellowship (NRF-NRFF13- 2021-0006), Singapore. Any opinions, findings, conclusions, or recommendations expressed in this material are those of the authors and do not reflect the views of the funding agencies.

\bibliographystyle{ACM-Reference-Format}
\bibliography{sample-base}
\clearpage
\appendix

\section{VLM Prompts}
\label{sec:prompt_descrip}
The prompt contains an instruction to generate a conversation,  the background information about the conversation, and a number of demonstration conversations. Example prompts for LIME, Grad-CAM, Integrated Gradients, and SHAP are shown in Figure \ref{fig:llm_prompt_lime}, \ref{fig:llm_prompt_grad_cam}, \ref{fig:llm_prompt_ig}, and \ref{fig:llm_prompt_shap} respectively. The input images are randomly selected from ImageNet and the explanations are generated by the corresponding XAI method.

\section{Measurement of Users’ Objective Understanding – Selection of Classification Models.}
\label{appendix_sec:objective}
The measurement aims to objectively measure participants' understanding of static explanations. We ask participants to choose, from three classification models, the most accurate on unobserved test data. 
All three classification models make the same decisions on 5 images, accompanied by static explanations from the same explanation method. The only differences between the three networks lie in their explanations. Hence, to make the correct selection, the participants must understand the explanations.

\hl{To select the images used in the study, we first sampled images from ImageNet and generated explanations for three classification models. Then, for each XAI method, we selected a subset of images where classification models with higher accuracy produced more reasonable explanations. Each XAI method has 16 images (15 for objective measure, 1 for conversation).}

Figure \ref{fig:q1_lime}, \ref{fig:q1_grad_cam}, \ref{fig:q1_ig}, \ref{fig:q1_shap} presents the full set of images for Grad-CAM, LIME, Integrated Gradients, and SHAP, respectively.

\section{Measurements of Users’ Subjective Perception}
\label{appendix_sec:subjective}
We also measure participants' subjective perception of static explanations, including their comprehension, acceptance, and trust. There are a total of 13 questions listed in Figure \ref{fig:q2_subjective}. All questions utilize a 7-point Likert scale for responses.

\iftrue
\begin{figure}
    \centering
    \includegraphics[width=\columnwidth]{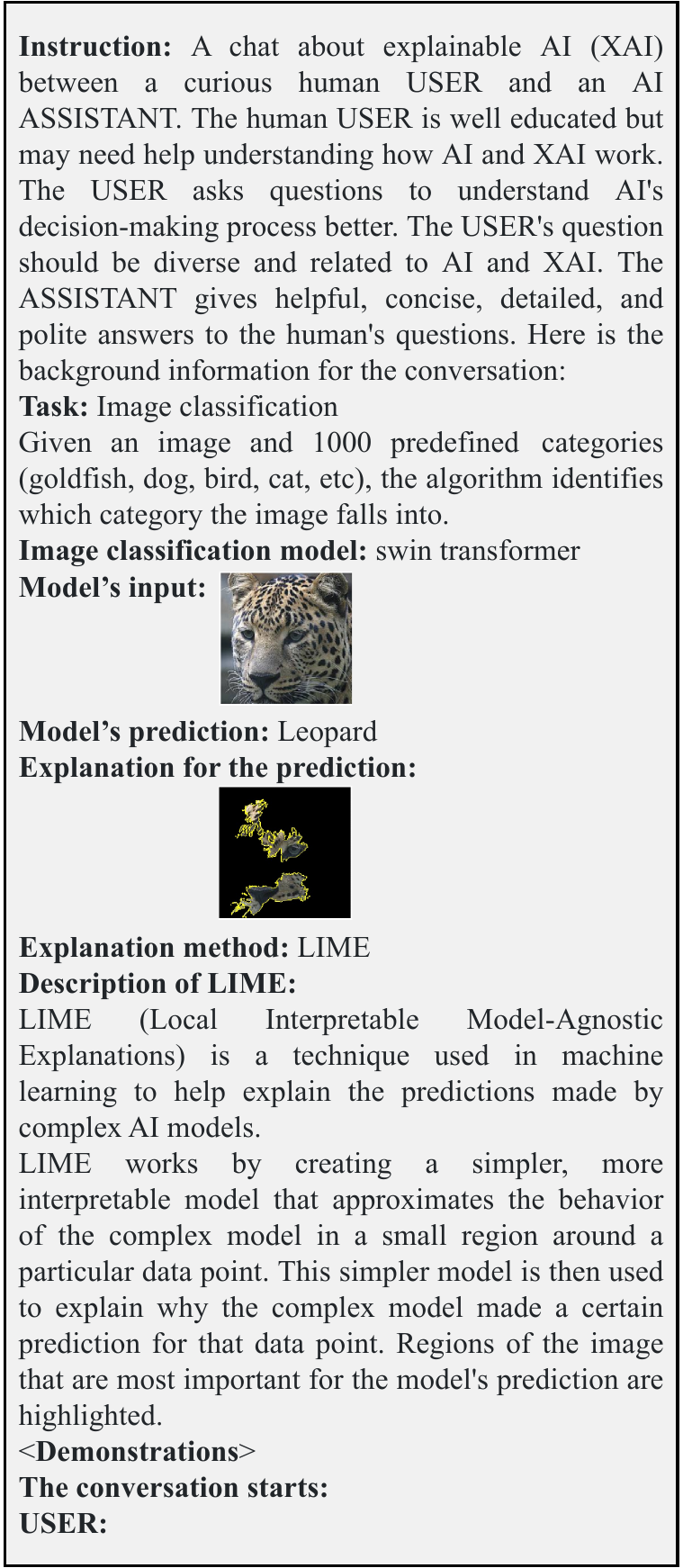}
    \caption{The VLM prompt about LIME.}
    \label{fig:llm_prompt_lime}
\end{figure}
\begin{figure}
    \centering
    \includegraphics[width=\columnwidth]{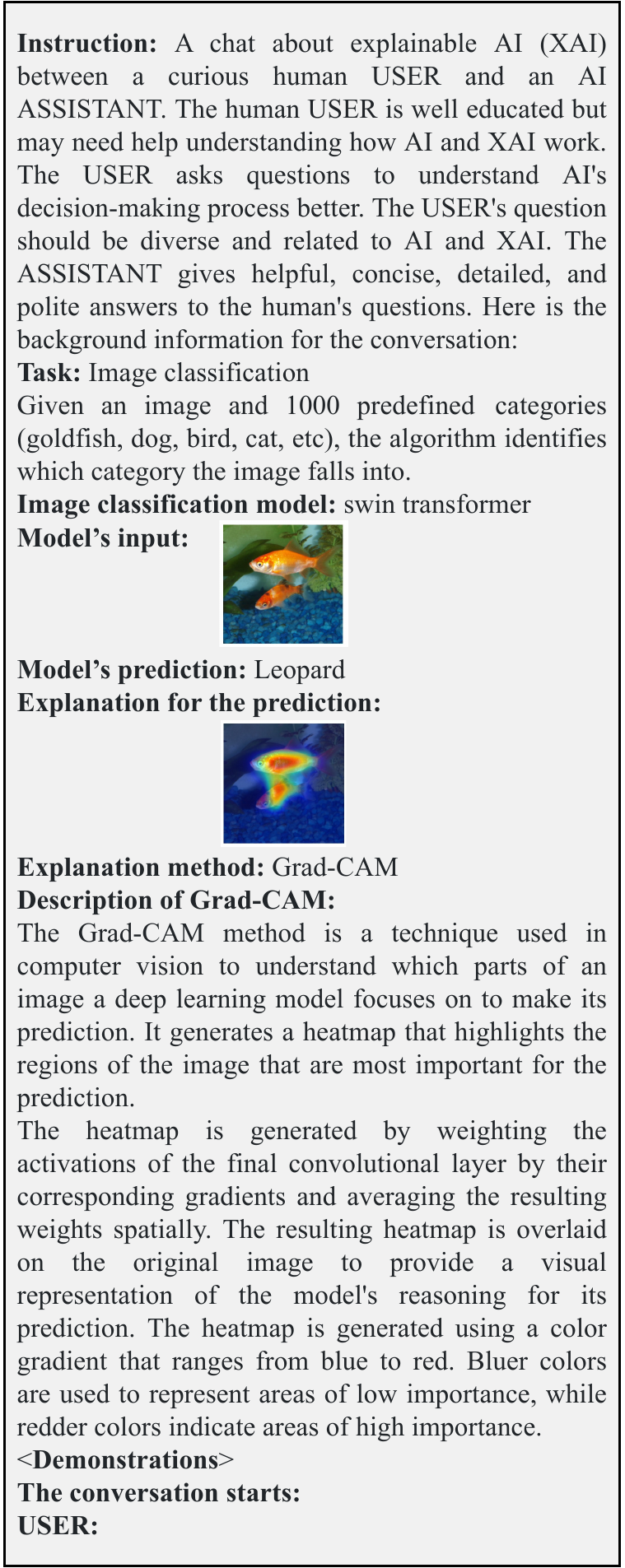}
    \caption{The VLM prompt about Grad-CAM.}
    \label{fig:llm_prompt_grad_cam}
\end{figure}
\begin{figure}
    \centering
    \includegraphics[width=\columnwidth]{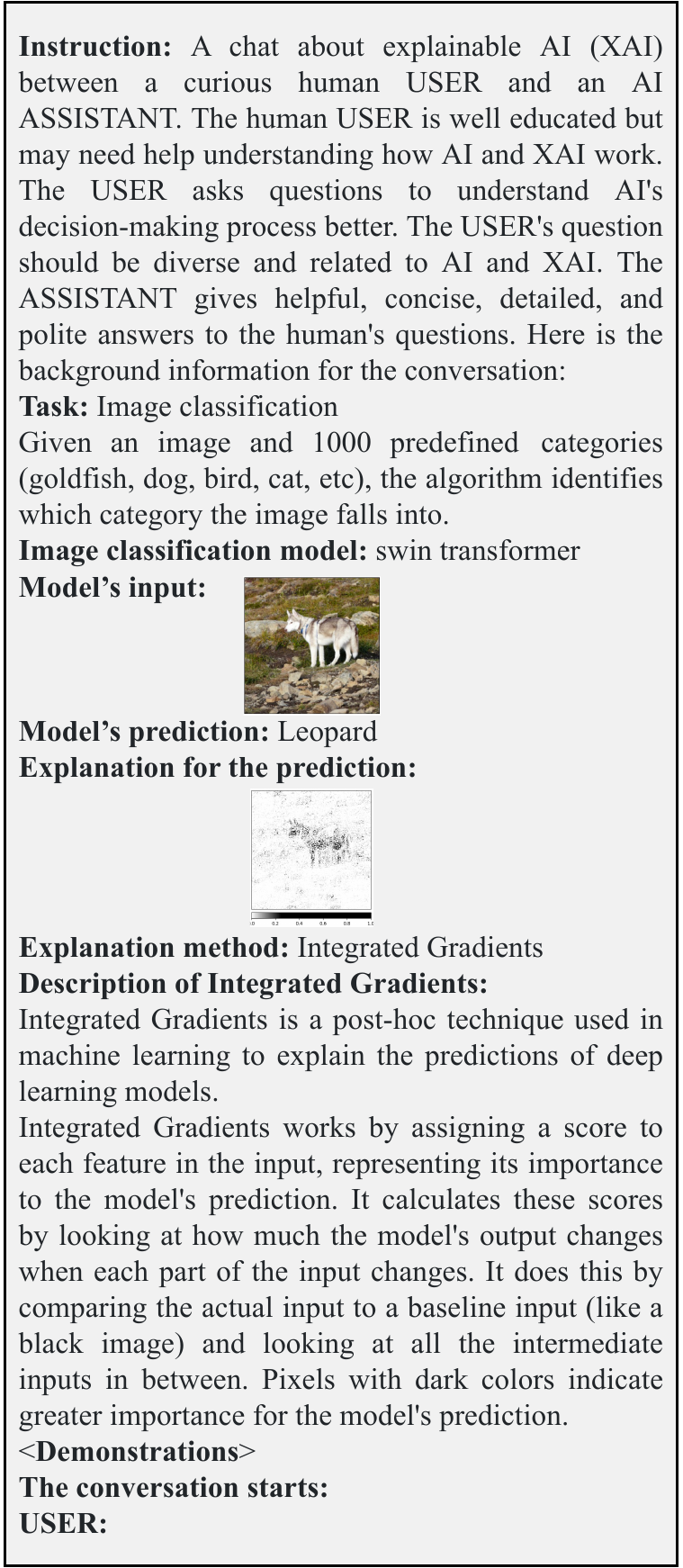}
    \caption{The VLM prompt about Integrated Gradients.}
    \label{fig:llm_prompt_ig}
\end{figure}
\begin{figure}
    \centering
    \includegraphics[width=\columnwidth]{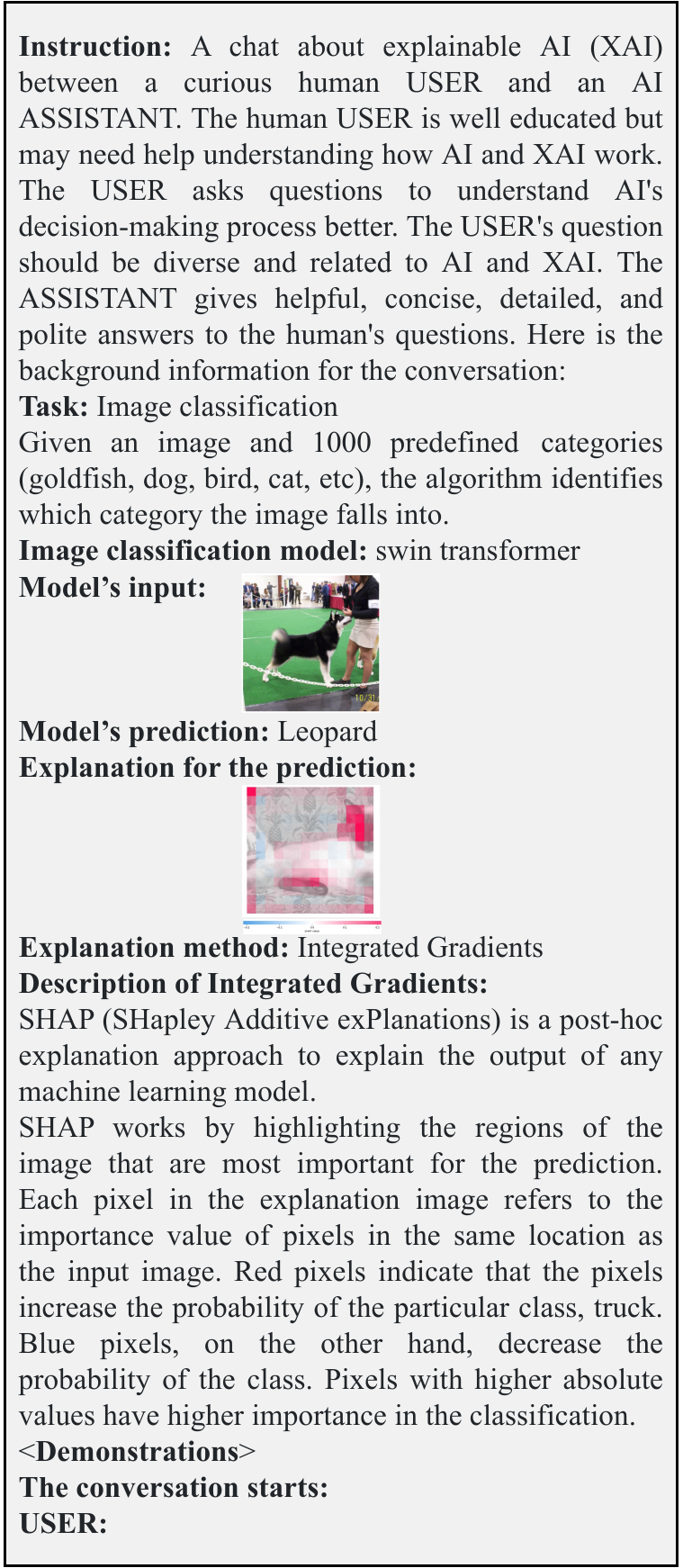}
    \caption{The VLM prompt about SHAP.}
    \label{fig:llm_prompt_shap}
\end{figure}

\begin{figure*}
    \centering
    \includegraphics[width=\textwidth]{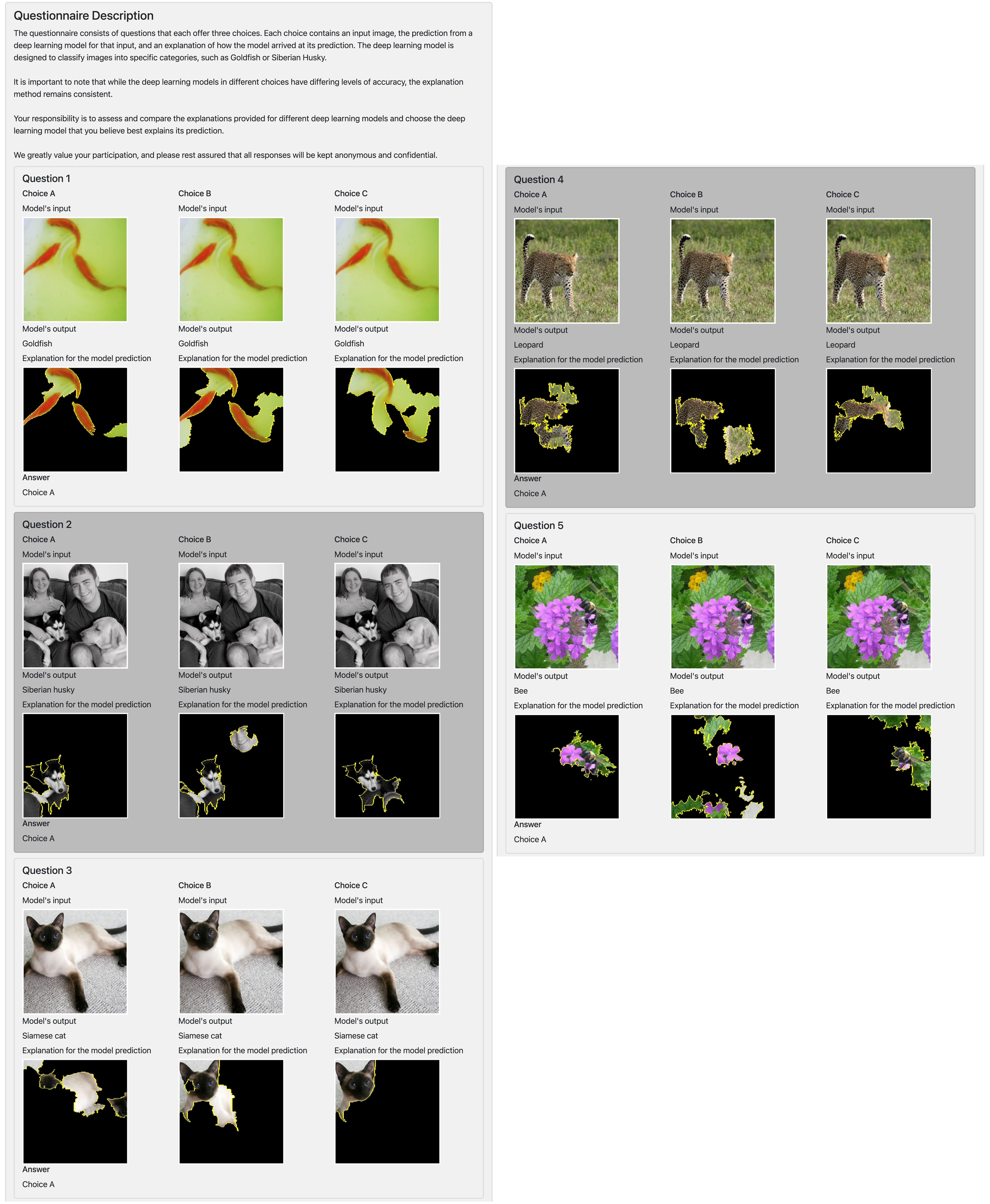}
    \caption{Questions used to measure participants' objective understanding of LIME.}
    \label{fig:q1_lime}
\end{figure*}
\begin{figure*}
    \centering
    \includegraphics[width=\linewidth]{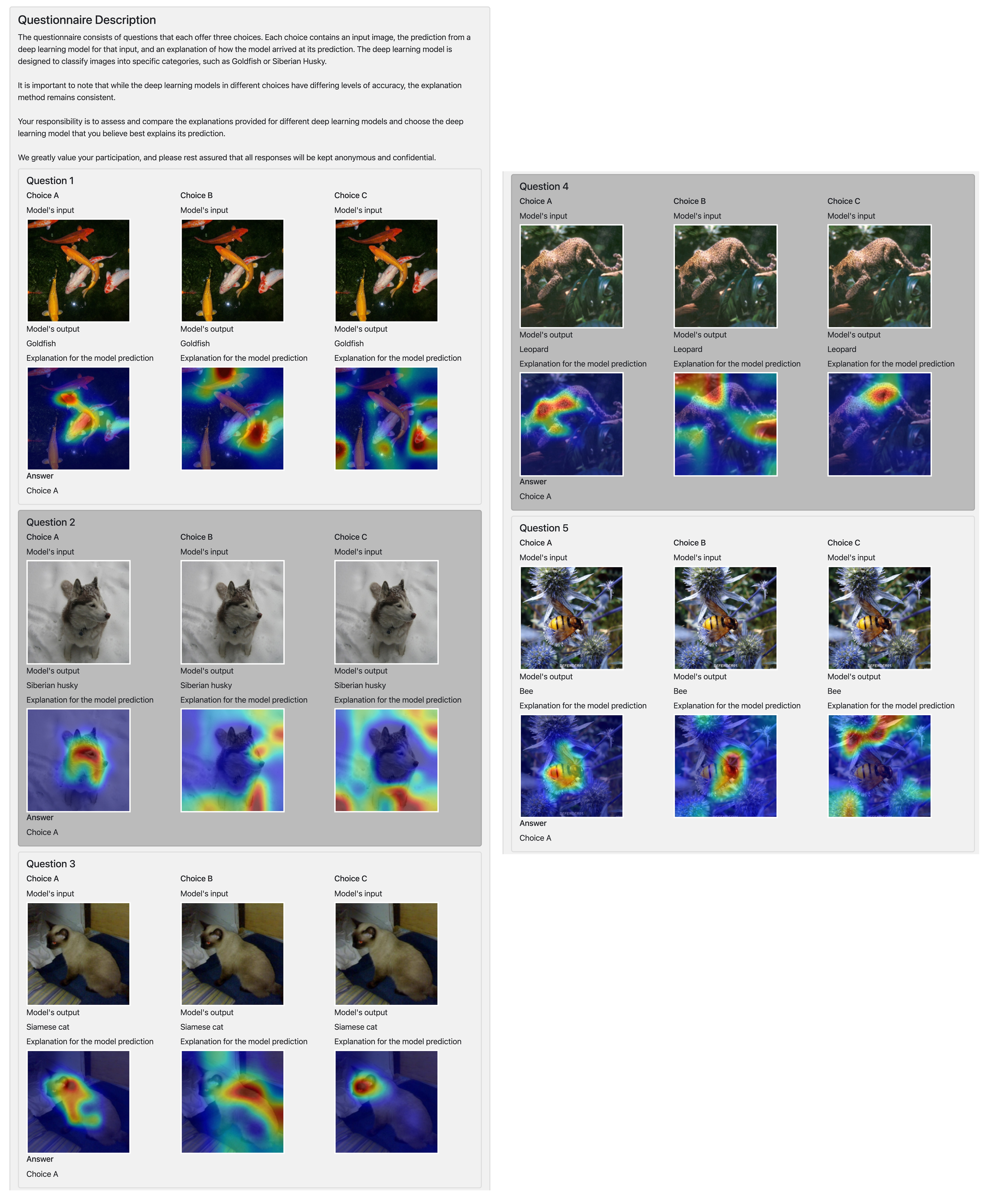}
    \caption{Questions used to measure participants' objective understanding of Grad-CAM.}
    \label{fig:q1_grad_cam}
\end{figure*}
\begin{figure*}
    \centering
    \includegraphics[width=\textwidth]{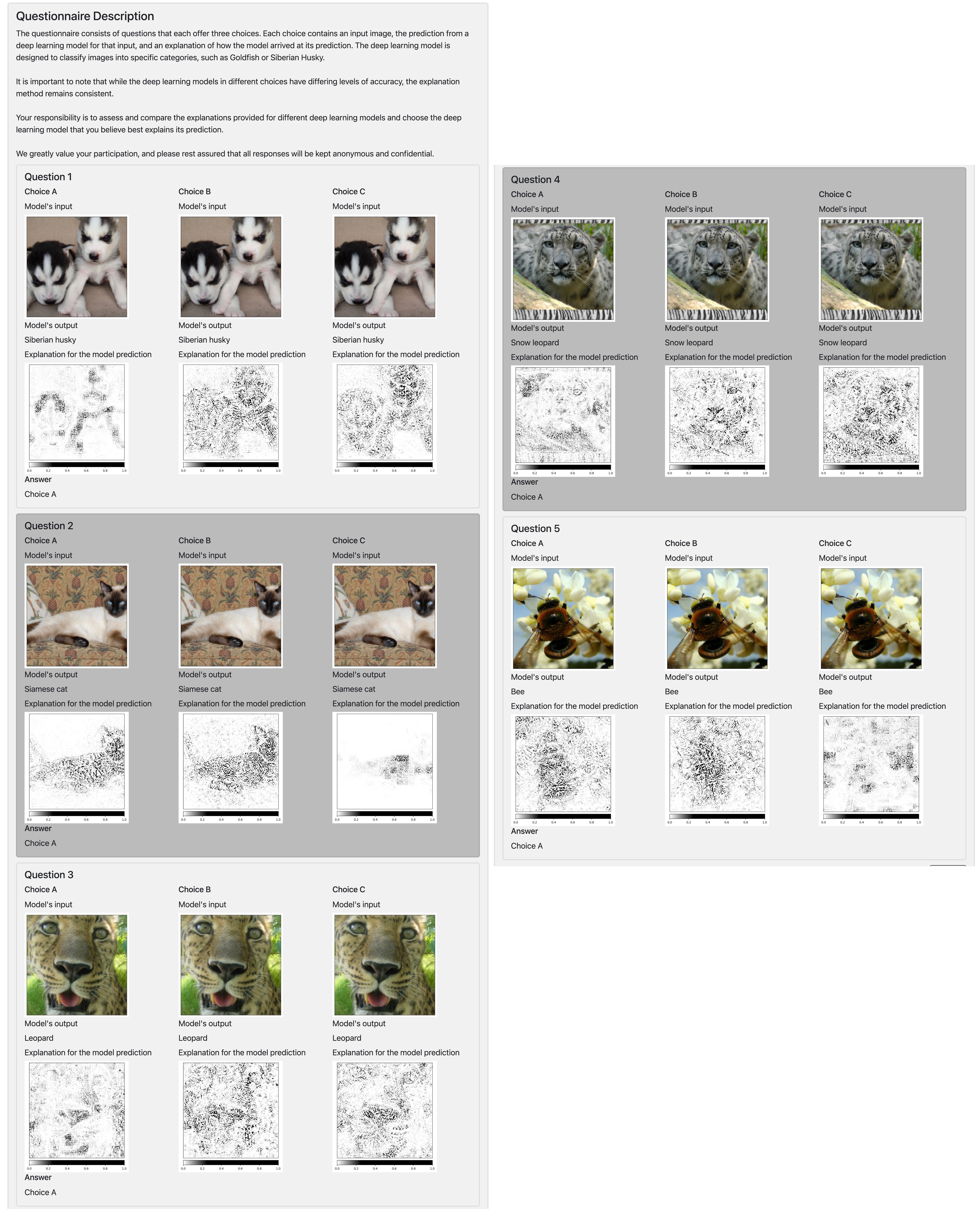}
    \caption{Questions used to measure participants' objective understanding of Integrated Gradients}
    \label{fig:q1_ig}
\end{figure*}
\begin{figure*}
    \centering
    \includegraphics[width=\textwidth]{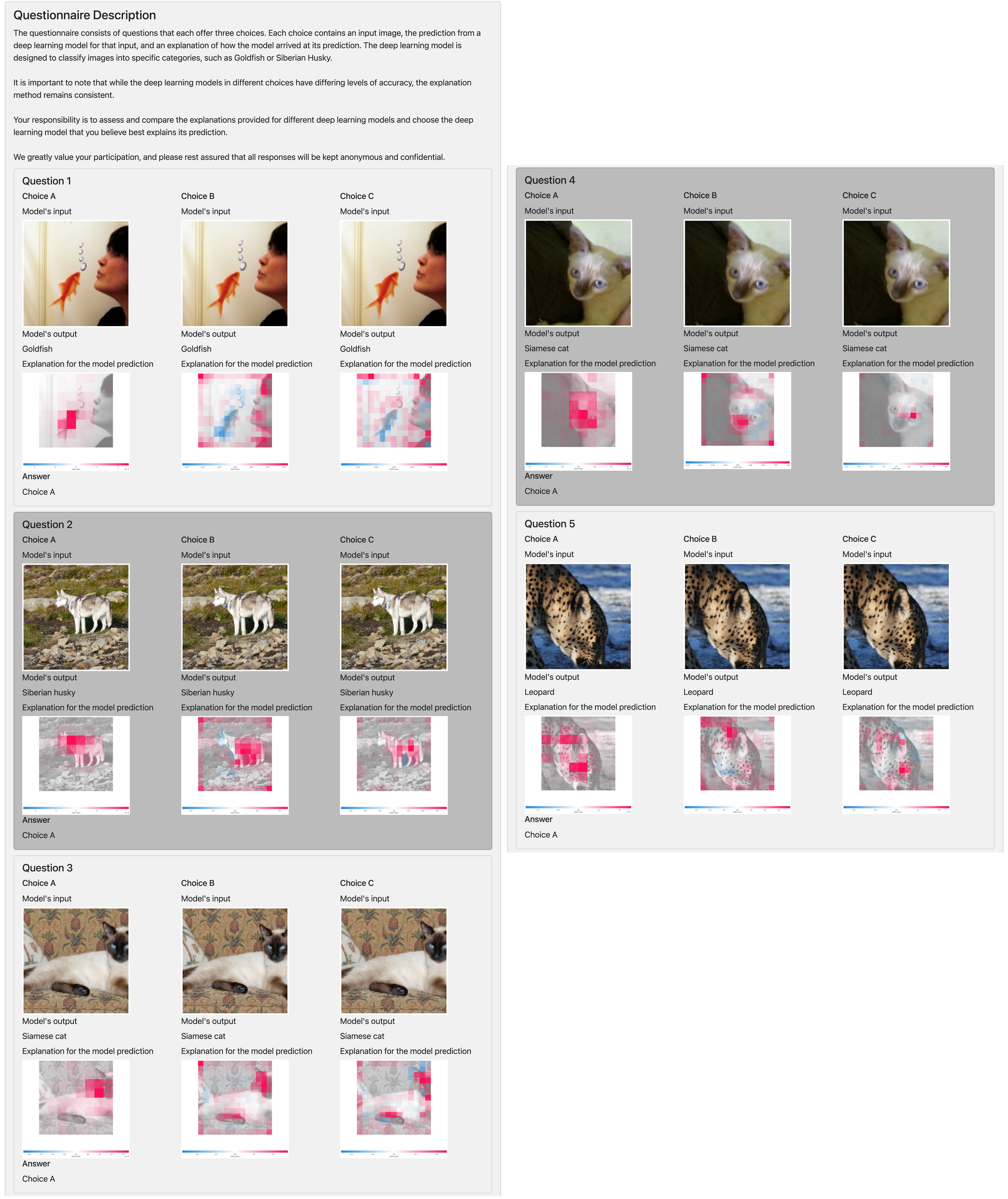}
    \caption{Questions used to measure participants' objective understanding of SHAP.}
    \label{fig:q1_shap}
\end{figure*}

\begin{figure*}
    \centering
    \includegraphics[width=\textwidth]{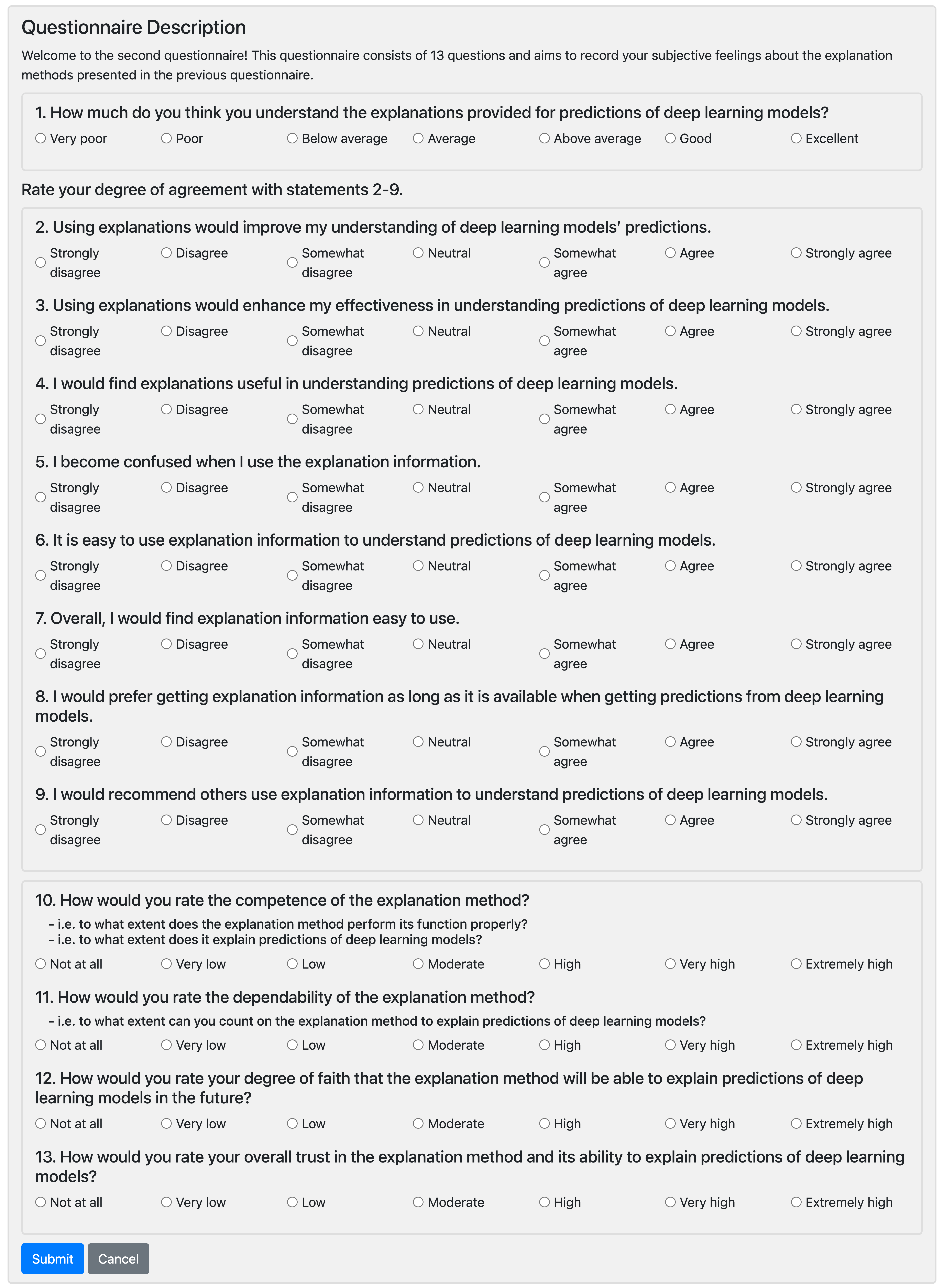}
    \caption{Questions users to measure participants' subjective perception of explanations. The user will respond to each question using a 7-point Likert scale.}
    \label{fig:q2_subjective}
\end{figure*}

\begin{figure}[!ht]
\centering\subfloat[LIME]{\includegraphics[width=0.49\linewidth]{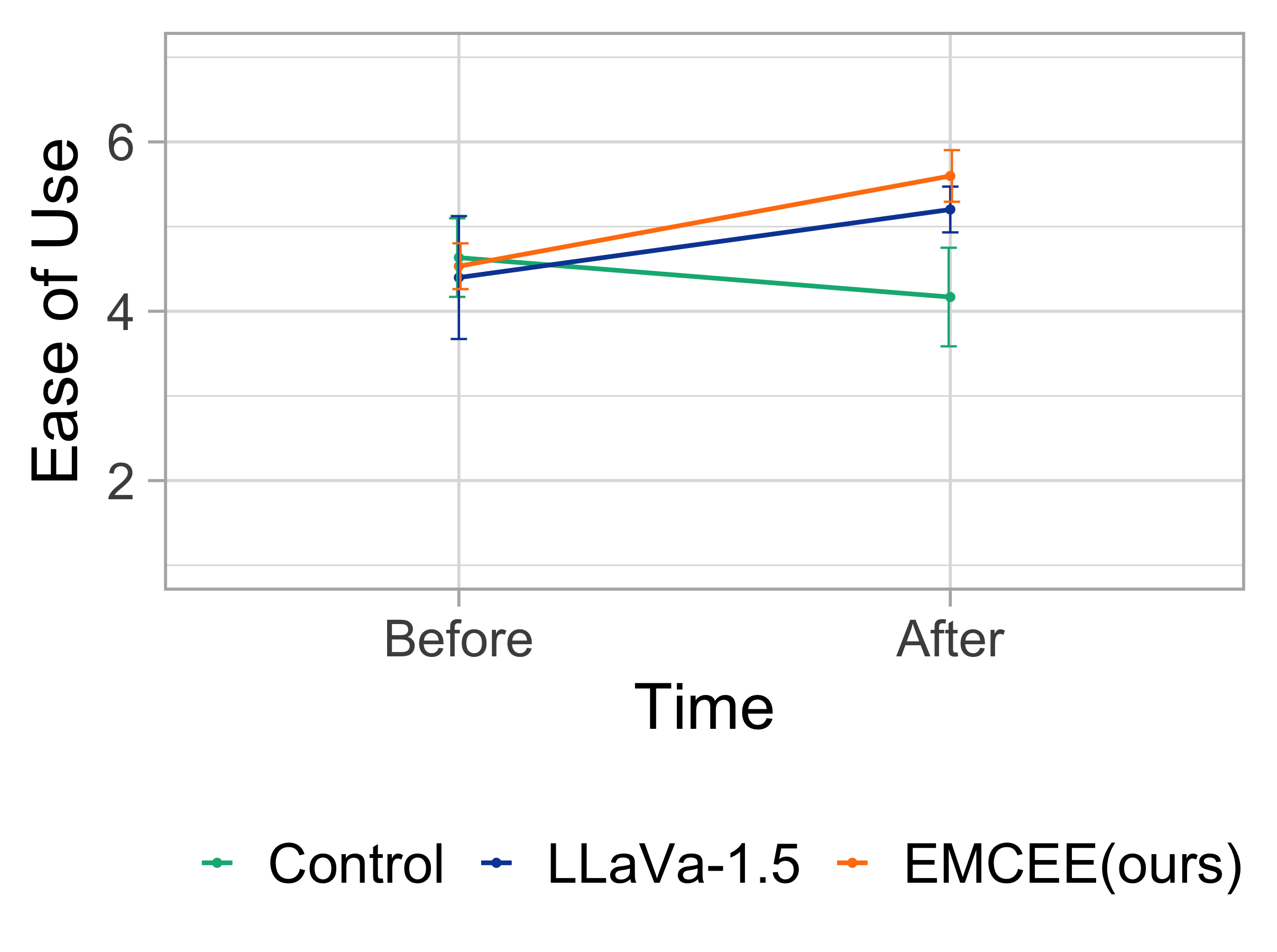}} 
\subfloat[Grad-CAM]{\includegraphics[width=0.49\linewidth]{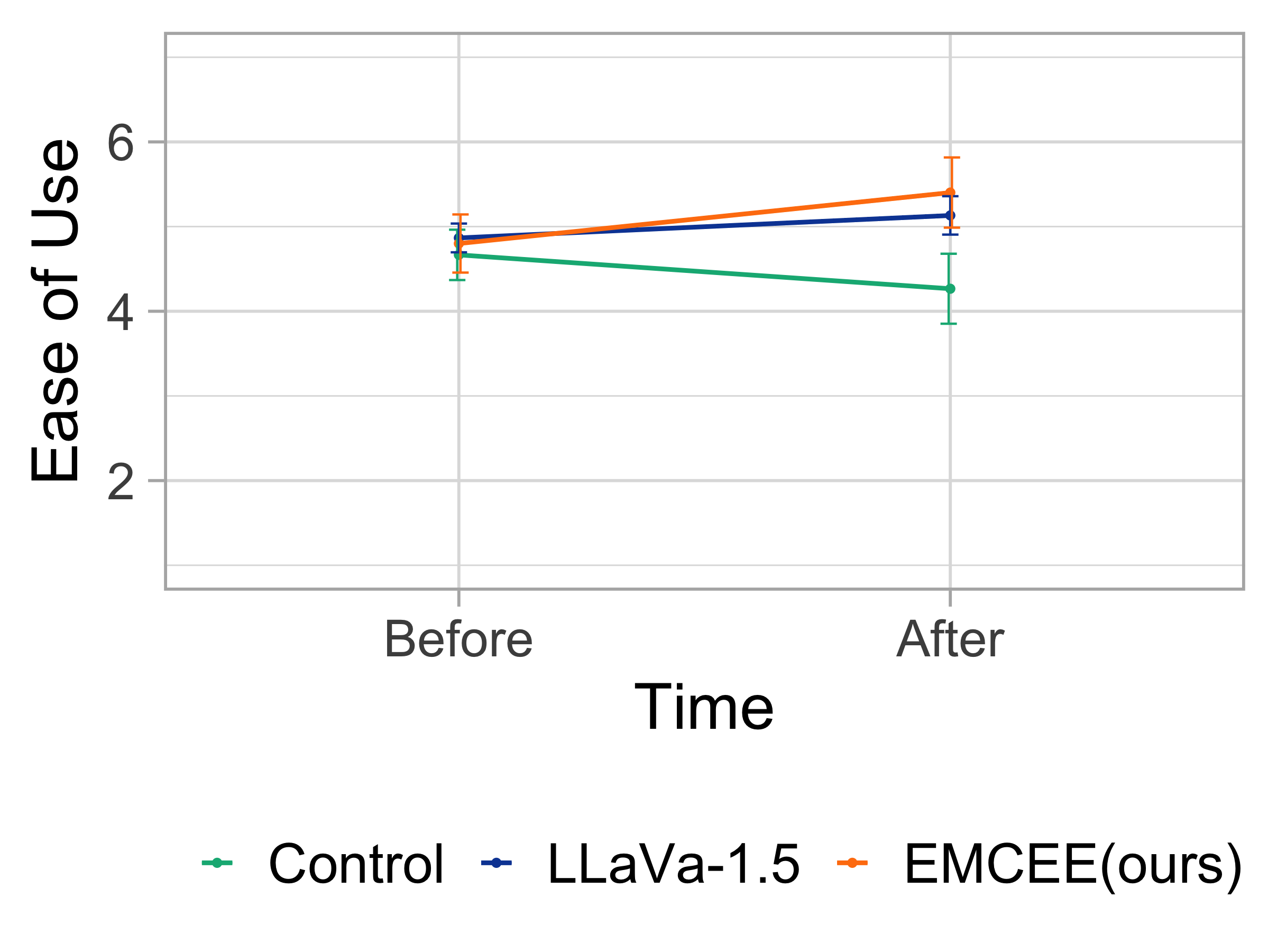}} \hspace{0.04\textwidth}
\subfloat[Integrated Gradients]{\includegraphics[width=0.49\linewidth]{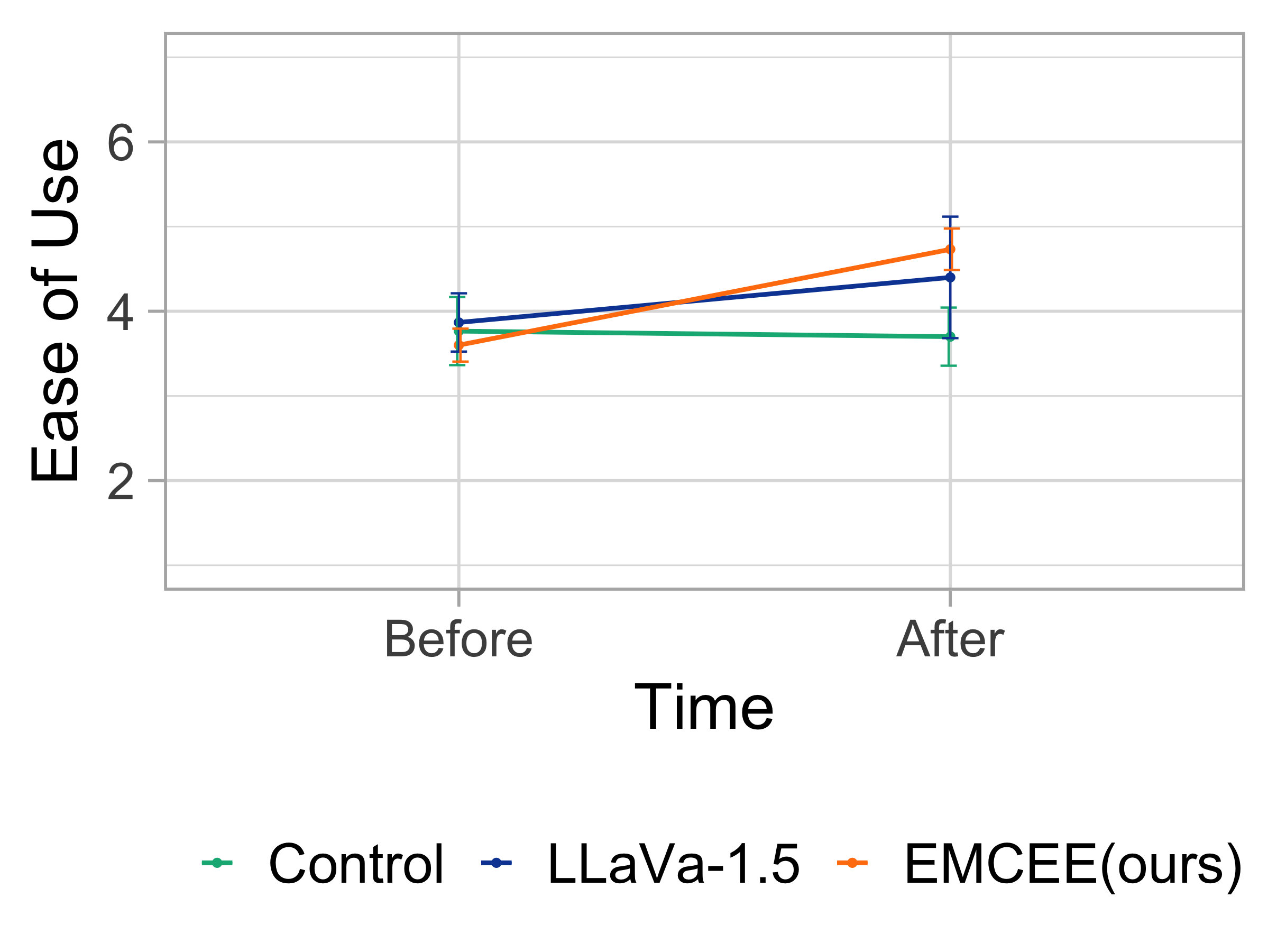}}
\subfloat[SHAP]{\includegraphics[width=0.49\linewidth]{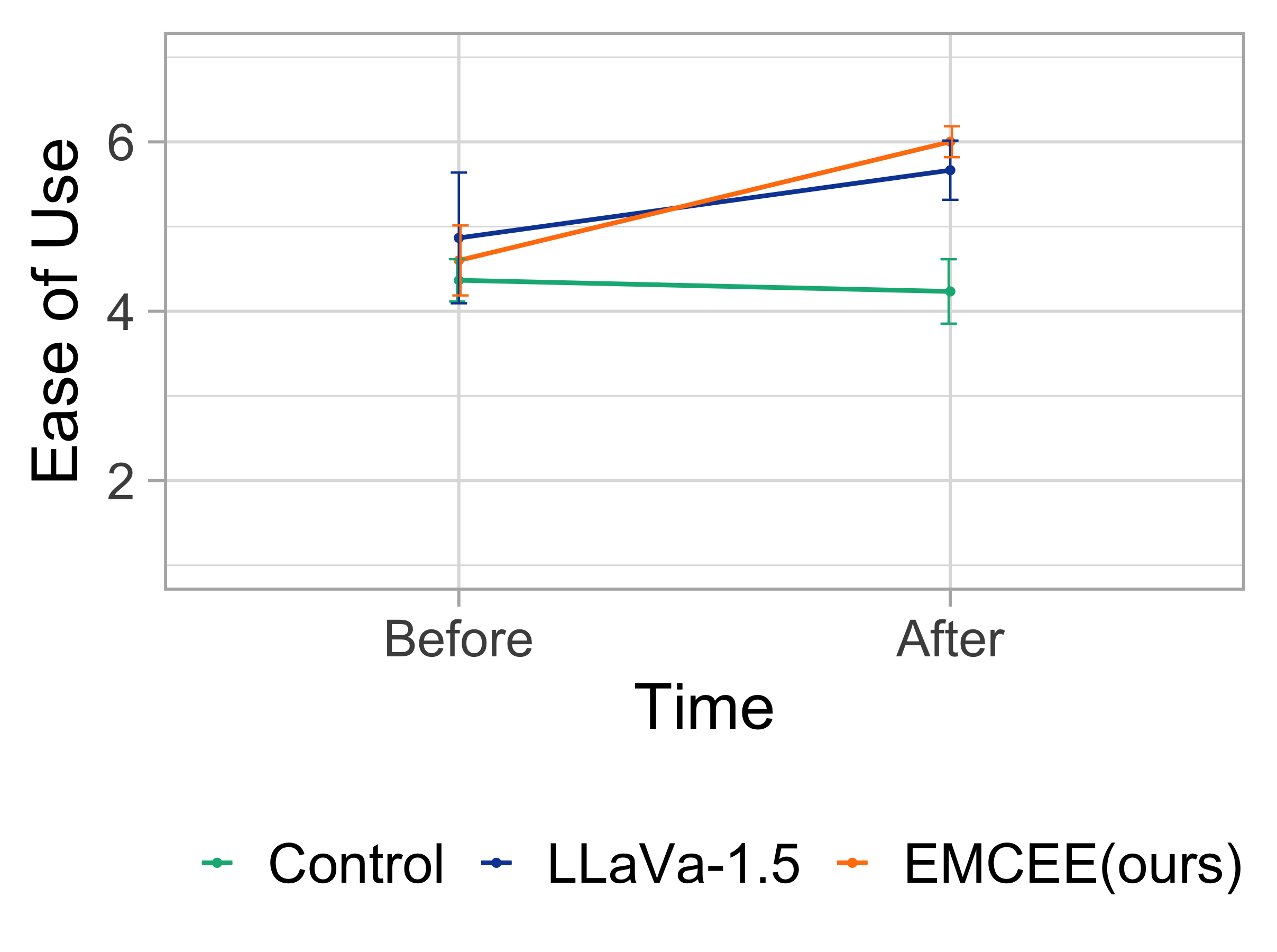}}
\caption{Participants' self-report ease of use score for (a) LIME and (b) Grad-CAM (c) Integrated Gradients (d) SHAP before and after conditions. }
\label{fig:eoe_interaction}
\end{figure}

\begin{figure}[!ht]
\centering\subfloat[LIME]{\includegraphics[width=0.49\linewidth]{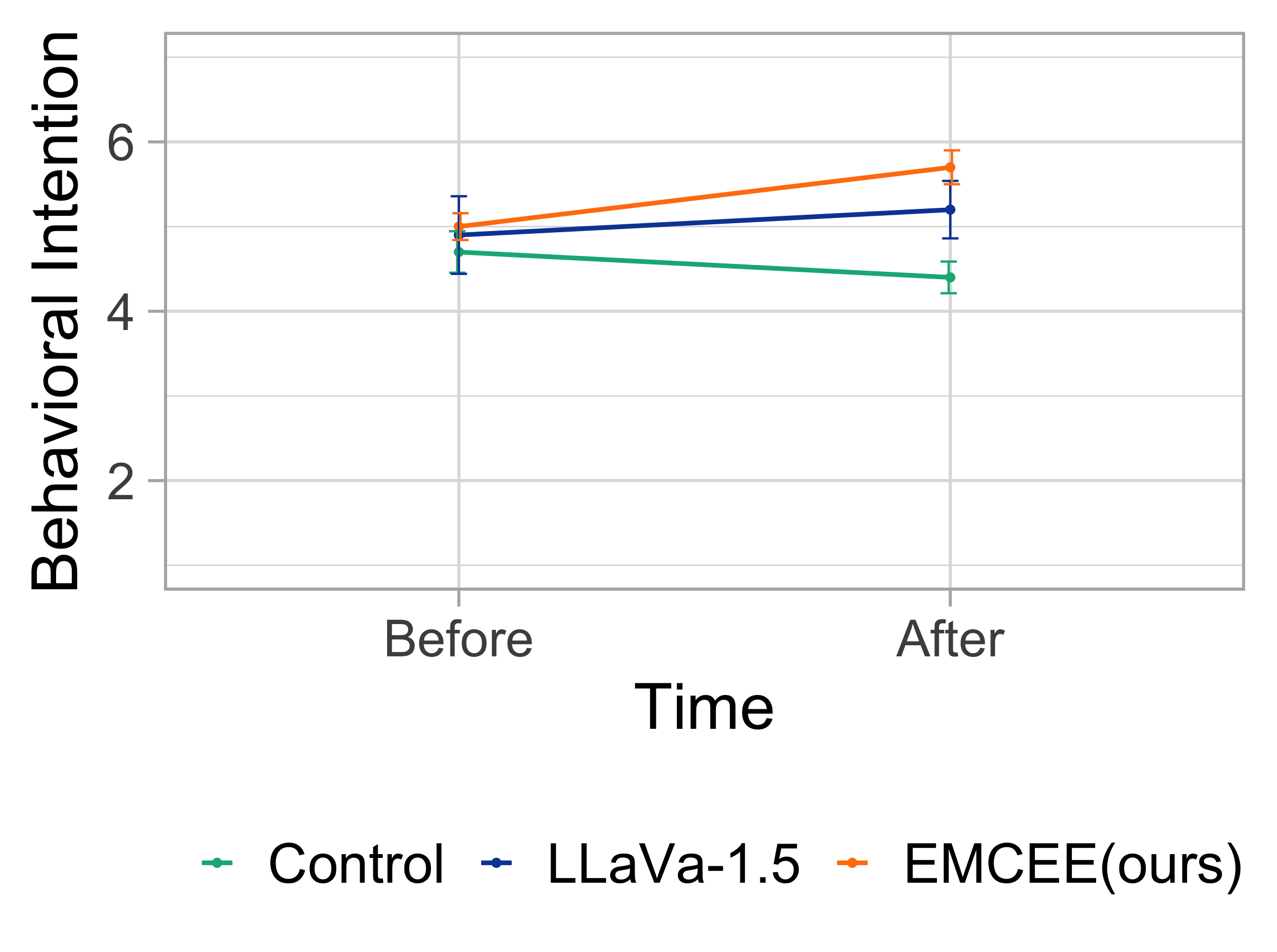}} 
\subfloat[Grad-CAM]{\includegraphics[width=0.49\linewidth]{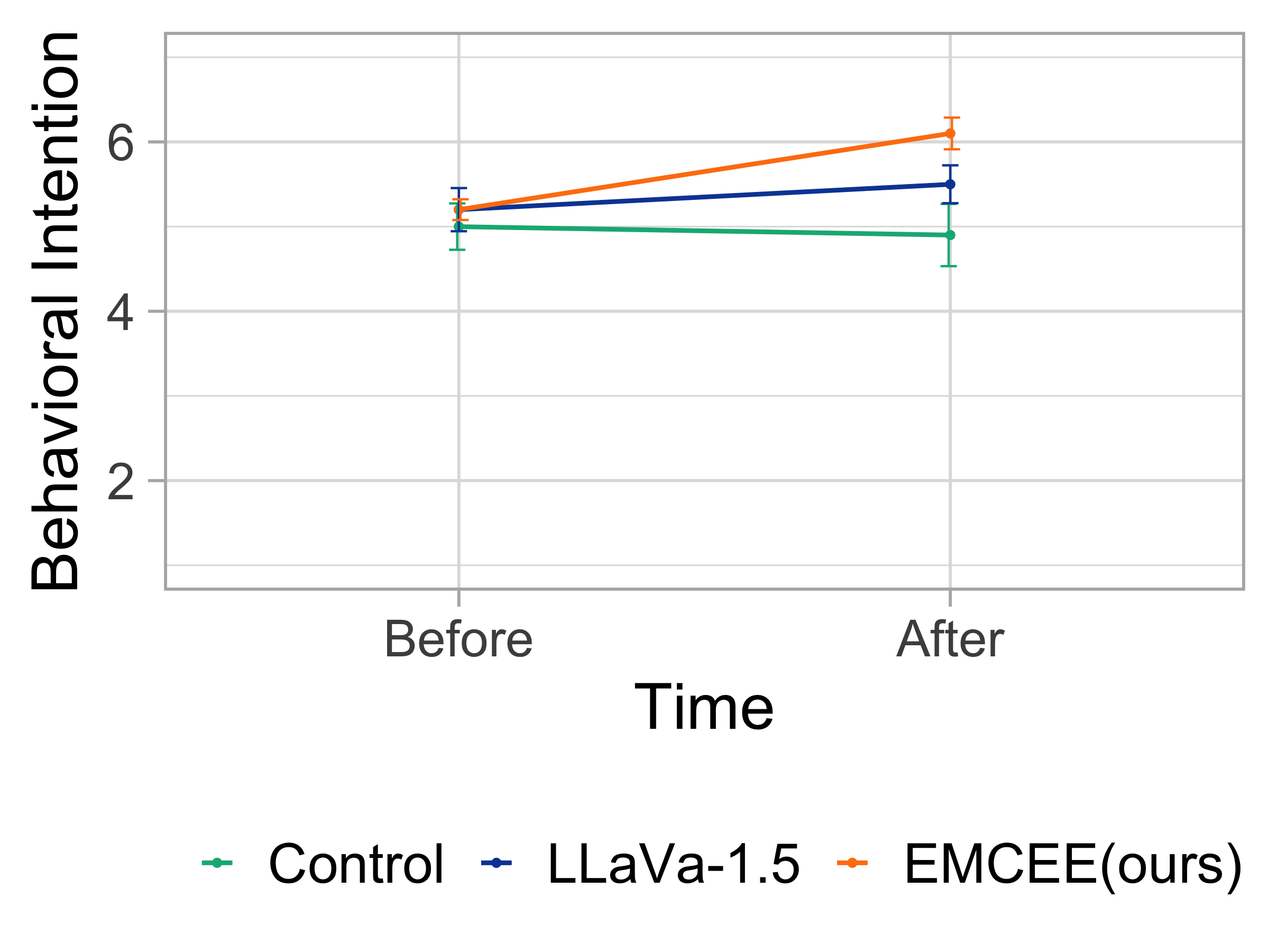}} \hspace{0.04\textwidth}
\subfloat[Integrated Gradients]{\includegraphics[width=0.49\linewidth]{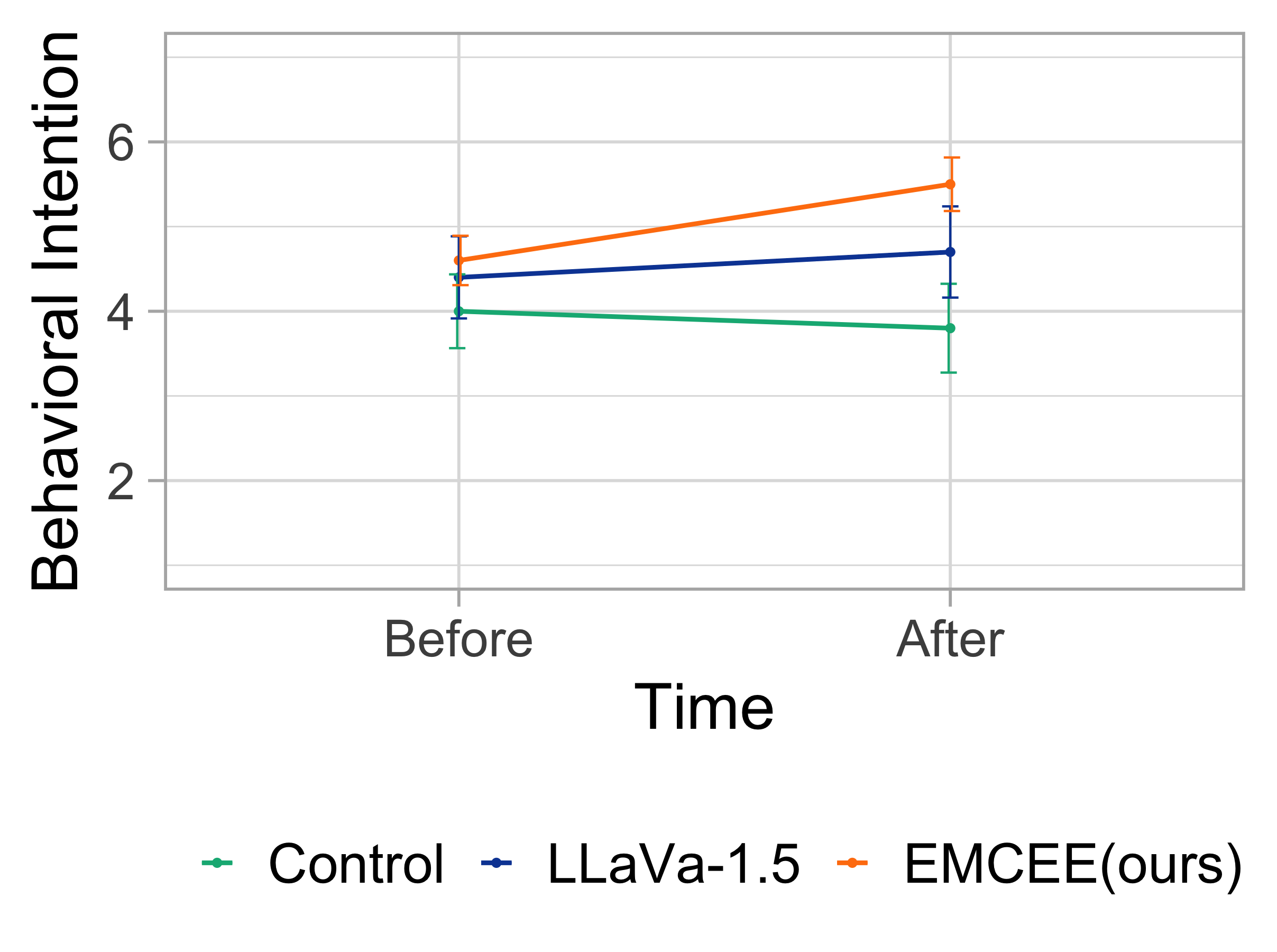}} 
\subfloat[SHAP]{\includegraphics[width=0.49\linewidth]{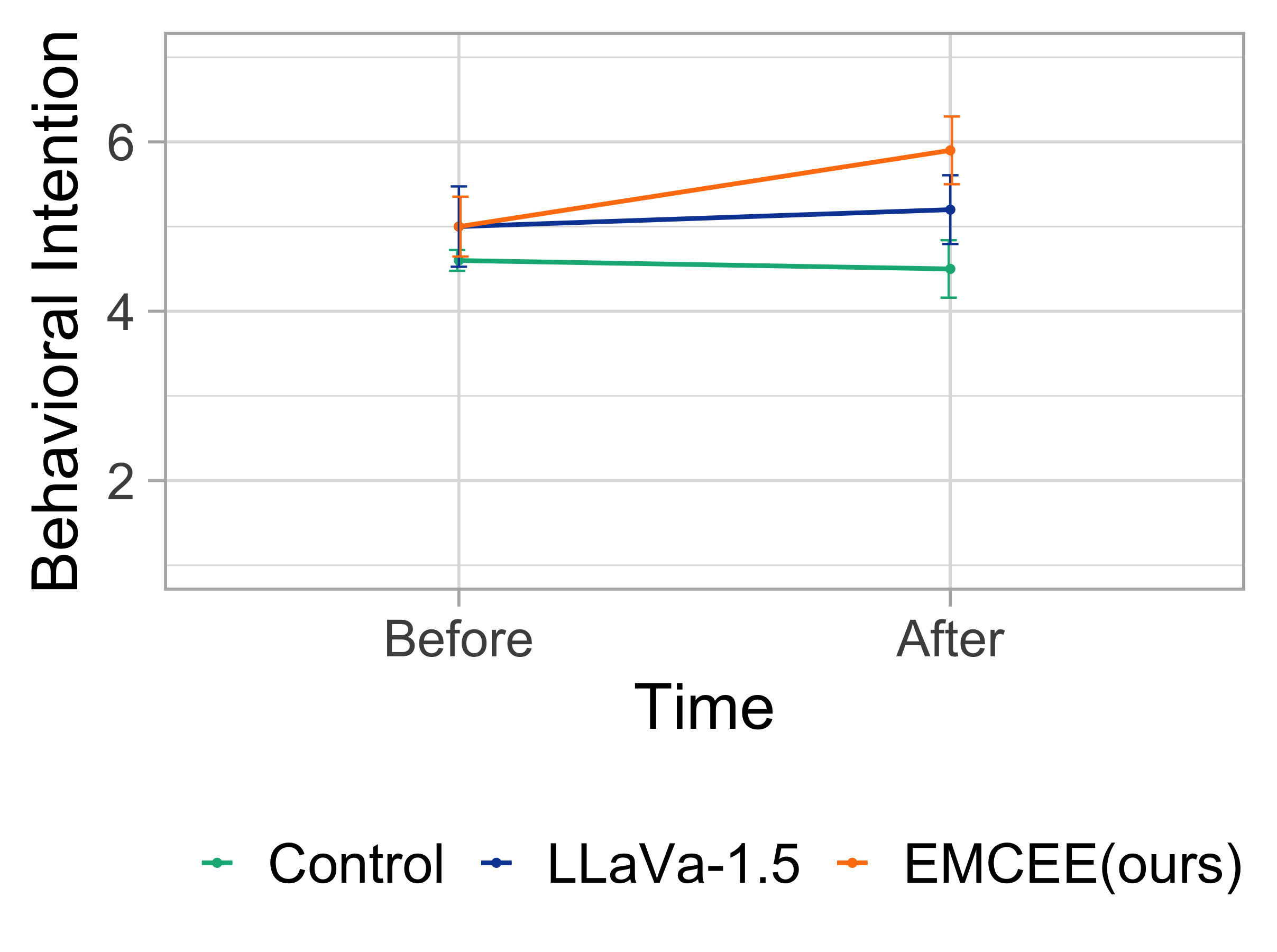}}
\caption{Participants' self-report behavioral intention score for (a) LIME and (b) Grad-CAM (c) Integrated Gradients (d) SHAP before and after conditions. }
\label{fig:bi_interaction}
\end{figure}

\fi

\end{document}